\documentclass[twocolumn,preprintnumbers,amsmath,amssymb,aps,prb,longbibliography]{revtex4-1}
\usepackage{graphicx}
\begin{document}

\title{Active Rheology and Anti-Commensuration Effects For Driven Probe Particles on Two Dimensional Periodic Pinning Substrates }
 
\author{C. J. O. Reichhardt and C. Reichhardt}
\affiliation
{
Theoretical Division and Center for Nonlinear Studies,
Los Alamos National Laboratory, Los Alamos, New Mexico 87545, USA}

\date{\today}

\begin{abstract}
For an assembly of particles interacting with a two dimensional periodic substrate, a series of commensuration effects can arise when the number of particles is an integer multiple of the number of substrate minima. Such commensuration effects can appear for vortices in type-II superconductors with periodic pinning or for colloidal particles on optical landscapes. Under bulk external driving, the pinning or drag on the particles is strongly enhanced at commensuration. Here we consider the active rheology of a {\it single} particle driven through an assembly of particles coupled to a periodic substrate at different commensurate conditions. For increasing density at fixed driving force, we observe nonmonotonic drag along with what we call an anti-commensuration effect where the drag or pinning effectiveness is reduced in commensurate states, opposite from the behavior typically observed under bulk driving. The velocity enhancement or drag reduction appears when the background particles form a crystalline state that is coupled more strongly to the substrate than to the driven particle, while under incommensurate conditions, the background particles are disordered and produce enhanced drag on the probe particle. The velocity noise of the driven particle has a narrow band signature at commensuration and a broad band signature away from commensuration. We map out the regions in which viscous flow, periodic flow, and a pinned phase appear. We show that the effects we observe are robust on both square and triangular substrate arrays and for both vortices in type-II superconductors and colloidal particles on optical landscapes.  
\end{abstract}

\maketitle

\section{Introduction}

A wide variety of systems can be described in terms of
a collection of interacting
particles coupled to a periodic substrate.
Commensuration effects arise
when the number of particles
is an integer or rational fractional multiple of the number of substrate minima
and
the system forms a highly ordered crystalline state
\cite{Bak82,Harada96,Hu97,Mangold03,Tung06,McDermott13,Vanossi13,Bohlein12,Reichhardt17}.
At incommensurate fillings, the particles can
remain in a lattice that floats above a weak substrate, or the
particle positions can be disordered by a stronger substrate; in each case,
the effectiveness of the pinning is reduced
\cite{Bak82,McDermott13,Vanossi13,Reichhardt17}. 
For fillings just outside of commensuration,
the system can be mostly 
ordered and contain
a small number of localized
excitations or solitons \cite{Bohlein12,Reichhardt17}.
Under an applied drive, at commensurate conditions the depinning threshold
above which motion occurs shifts to higher drives,
while for incommensurate conditions,
the depinning threshold is depressed or can show
distinct steps due to the separate depinning of the solitons and the
bulk particles;
additionally, if the particles are already moving,
the velocity is strongly suppressed at
commensuration and
is largest for incommensurate conditions
\cite{Bohlein12,Reichhardt17,Gutierrez09,Avci10,Vanossi12}.
The ordering of the 
particles at commensuration
depends on the dimensionality and  geometry of the periodic substrate.
For a two-dimensional (2D) particle assembly on a quasi-one-dimensional (q1D) substrate,
smooth transitions occur between ordered commensurate and disordered
incommensurate states, such as crystal and smectic states
\cite{Daldini74,Dobrovolskiy12,LeThien16},
while for a 2D system with
a 2D periodic substrate,
the commensurate effects can be sharper and 
consist of transitions
between various types of 2D crystals
\cite{Harada96,Mangold03,Tung06,McDermott13,Bohlein12,Reichhardt17,Baert95,Reichhardt98a,Reichhardt02,Brunner02}. 
Examples of 2D particle systems coupled to a 2D periodic substrate
include vortices in type-II superconductors with periodic pinning,
\cite{Harada96,Reichhardt17,Gutierrez09,Avci10,Baert95,Reichhardt98a,Martin97,Grigorenko03,Berdiyorov06,Latimer13,Reichhardt01a,Trastoy14,Sadovskyy17,Wang18},
vortices in Bose-Einstein condensates with optical traps \cite{Tung06},
colloidal particles on optical traps
\cite{Bohlein12,Reichhardt02,Brunner02,Reichhardt02a,Brunner02,Brazda18} or
etched surfaces \cite{McDermott13,Vanossi12,OrtizAmbriz16,Stoop20},
active matter on patterned substrates \cite{Reichhardt21},
cold atom systems \cite{Lewenstein07},
dusty plasmas on 2D arrays \cite{Huang21}, 
skyrmions in nanostructured samples \cite{Reichhardt15a,Feilhauer20},
coupled nanomagnetic islands \cite{Nisoli13},
and numerous types of atomic or frictional systems
\cite{Vanossi13,Hasnain14,Koplik10,Cam21}. 
In many of these systems, when driving is applied, dynamic transitions can produce
signatures in the velocity-force curves
\cite{Vanossi13,Bohlein12,Reichhardt17,Gutierrez09,Avci10,Vanossi12,Reichhardt01a,Hasnain14,Cam21,McDermott13a}
and  structural changes can occur in the particle positions
at the depinning threshold or for higher drives
\cite{Reichhardt17,Vanossi12,Reichhardt01a,Stoop20,Reichhardt98,McDermott13a}.
The effect of varied commensuration conditions on pinning has been
heavily studied for
vortices in type-II superconductors,
where the ratio of the number of particles to the number of pinning sites can
be varied easily by changing the magnetic field, and where
peaks in the critical currents
or reduced vortex velocities
appear at the matching fields
\cite{Reichhardt17,Gutierrez09,Avci10,Baert95,Reichhardt98a,Martin97,Berdiyorov06,Trastoy14,Durkin16,Sadovskyy17,Wang18}
and at fractional matching fillings
\cite{Grigorenko03,Berdiyorov06,Latimer13,Reichhardt01a}.
Peaks in the depinning force at commensurate fillings have also been studied for
colloidal systems \cite{McDermott13,Bohlein12,Vanossi12}.

In studies of depinning effects
at commensurate and incommensurate fields
in the systems above,
the
driving was applied to all of the particles at the same time; however,
in many systems it is possible to drive only a single individual particle
through the sample.
Measurement of the dynamics of a single
driven probe particle though a medium of other background particles
is known as active rheology
\cite{Hastings03,Habdas04,Squires05,Voigtmann13,Zia18,Gazuz09,Reichhardt21a}.
Here, an individual particle is dragged under different
conditions such as constant force, constant  velocity, or increasing
force, and the resulting pinning threshold or
viscous drag can be measured as the system parameters are varied.
In colloidal assemblies, active rheology has been applied
to glassy systems
\cite{Habdas04,Gazuz09,Winter12,Senbil19,Yu20}, where there can be
an onset of a finite depinning threshold or a rapid increase in the
drag as the glass transition is approached.
For granular systems,
investigations have revealed how the drag, pinning,
and fluctuations on an individual driven grain change
as the jamming transition is approached, along with 
the range out to which the surrounding particles
become disordered as the driven particle moves through the sample
\cite{Drocco05,Candelier10,Kolb13}.
Other studies have revealed the changes in the motion
\cite{Dullens11} or the viscoelastic response
\cite{AbaurreaVelasco20} of the driven particle 
as a transition occurs from elastic to plastic distortions in the surrounding
medium.
For systems such as superconducting vortices or magnetic skyrmions, individual 
driven particles can interact  both with the other particles and also with
a substrate in the form of a pinning
landscape \cite{Straver08,Reichhardt09a,Auslaender09,Hanneken16,Casiraghi19}.
The most studied example of this
involves the dragging of individual vortices in order
to examine vortex-pinning 
and vortex-vortex
interactions
\cite{Straver08,Reichhardt09a,Auslaender09,Shapira15,Veshchunov16,Kremen16,Polshyn19}.
The depinning threshold for the individually driven vortex can be nonzero even
in the absence of random pinning due to the elasticity of the surrounding
vortex lattice, while when random pinning is added to the system, the
depinning threshold is enhanced
\cite{Reichhardt08}. 

For individually dragged vortices moving
over periodic pinning arrays containing pinned vortices,
stick-slip motion can appear,
and the dynamics of the driven vortex can vary depending on the
orientation of the driving direction
with respect to
the symmetry of the underlying pinning array \cite{Ma18}.
It has also been proposed  that for individually driven vortices on 
a periodic pinning array
in a system which hosts Majorana fermions,
it is possible to take advantage of the different dynamics
to create various types of quantum logic gates \cite{Ma20}.
An open question is how the dynamics or
drag on a driven particle such as
a superconducting vortex
changes as the system is tuned from an incommensurate to a commensurate
state. One
possibility is that the drag will
be enhanced at matching, as found in bulk driven systems. 

Although active rheology has been studied extensively for 
systems in the absence of a substrate,
far less is known about when happens when 
a periodic substrate is present. 
Here we examine superconducting vortices and colloidal particles
interacting with 2D square and triangular pinning arrays, where
we drive a single particle through the medium
for varied filling fractions and driving forces.
At a constant
driving  force and in the absence of pinning,
the mobility of the probe particle decreases monotonically with increasing
system density due to increased collisions
with the background particles; however, when the periodic pinning array 
is present, a series of velocity peaks appear at commensurate fillings
when the system forms an ordered lattice.
We call these peaks anti-commenensuration effects 
since for bulk driving in most systems with pinning,
the velocity of the particles is reduced at the matching conditions.
For our single driven particle, the velocity
enhancement arises when the surrounding particles become more strongly
coupled to the substrate than to the probe particle under a matching condition.
At incommensurate fields where the system is disordered, the probe particle
interacts more strongly with the background particles and experiences a higher
effective drag force.
At a matching condition, the velocity of the probe particle
exhibits a periodic or narrow band noise signature,
while at incommensurate fillings,
the probe particle motion is disordered or chaotic,
giving rise to a broad band noise
signal.
We also find that the pinning effect on the probe particle is reduced at the matching conditions. 
As function of pinning force and filling fraction,
we map out the pinned, viscous flow, and ordered flow phases. 
In general, under a constant drive the probe 
particle moves at a higher velocity in the presence of pinning than in the absence of 
pinning since the background particles cannot be entrained by the
probe particle when they are trapped by pinning sites.
In the case of colloidal particles, as the effective charge
of the particles is reduced, we observe a transition
from a pinned state to an unpinned state. This is the opposite
of the behavior found in bulk driven systems, where a weakening of
the colloid-colloid interactions causes the colloids to become
more strongly coupled to the substrate. 

\section{Simulation and System}

We consider a two-dimensional system containing $N$ particles along with
a periodic array of $N_p$ pinning sites in the form of
potential energy minima that can each capture a single particle.
A matching condition occurs when
the ratio $f=N/N_{p}$ is an integer,
and it is known from previous studies of superconducting vortices and 
colloidal particles that at such fillings,
the particles form an ordered crystal state
\cite{Harada96,Tung06,McDermott13,Bohlein12,Reichhardt17,Vanossi12,Baert95,Reichhardt98a,Reichhardt02}. At incommensurate fillings, the system
is disordered or contains localized regions of excitations.
It is also possible for ordered crystalline states to appear at certain
fractional matching conditions
such as $f=1/2$ on a square
pinning array or $f=1/3$ on a triangular pinning array
\cite{Grigorenko03,Berdiyorov06,Reichhardt01a}. 
In the first part of the work, we focus on vortices in type-II superconductors
with periodic pinning, 
where there has been extensive numerical modeling
\cite{Reichhardt17,Reichhardt98a,Berdiyorov06,Reichhardt01a,Reichhardt98,Ma18,Ma20}
and experimental studies
\cite{Harada96,Reichhardt17,Baert95,Martin97,Grigorenko03}
examining the commensuration effects under bulk driving.
Here we instead drive only a single vortex.
The equation of motion for vortex $i$ is given by
\begin{equation}
\alpha_d {\bf v}_i   = {\bf F}_i^{vv} + {\bf F}_i^{p} + {\bf F}^{D}_i.
\end{equation}
Here ${\bf v}_i=d{\bf r}_i/dt$ is the velocity 
and ${\bf r}_i$ is the position of vortex $i$.
The dynamics is overdamped and we set the damping constant
$\alpha_d$
equal to unity.
Vortex-vortex interactions are described by the term
${\bf F}_i^{vv} = \sum_{j = 1}^{N} K_1(r_{ij}) \hat{\bf{r}}_{ij}$,
where $r_{ij} = |{\bf r}_i - {\bf r}_j|$, $\hat{{\bf r}}_{ij} = ({\bf r}_i - {\bf r}_j)/r_{ij} $,
and $K_1$ is the
modified Bessel function which decays exponentially for large $r$.
Pinning sites
are modeled as
parabolic traps with a maximum range of $r_{p}$
that produce a pinning force described by
${\bf F}_i^{p} = \sum_{k = 1}^{N_p} (F_p/r_p) ({\bf r}_i - {\bf r}_k^{(p)}) \Theta( r_p - |{\bf r}_i - {\bf r}_k^{(p)}| )$.
Here $F_{p}$
is the maximum pinning force and $\Theta$ is the Heaviside step
function.
A driving force ${\bf F}^{D} = F_{D} \hat{\bf{x}}$ is applied to only one
vortex with $i=D$
and is always aligned with the $x$ direction; the driving force on all of the
other vortices with $i \neq D$ is set to zero.
For the driven or probe vortex, we measure the net velocity
$V={\bf v}_{i=D} \cdot {\bf \hat{x}}$.
At a given driving force,
we consider both time series data
and the time averaged velocity $\langle V\rangle$.
In some cases, we apply a constant drive, while in other cases
we sweep the driving force from zero to a maximum value in order to
measure the depinning force $F_{c}$ and the effective drag. 
The initial positions of the vortices are obtained
using simulated annealing \cite{Reichhardt98a},
where the sample is initialized at a high temperature using random Langevin
kicks and gradually cooled to zero in order to obtain a low energy state.
This process is repeated to obtain
starting configurations for different filling factors $f$. 
In superconducting  
systems,
the vortex density is proportional to the magnetic field $B$
and the filling factor is 
measured relative to the $1:1$ vortex to pinning site matching condition
known as the matching field $B_{\phi}$,
such that $f = B/B_{\phi}$. 
The sample
is of size $L \times L$ with $L = 36\lambda$,
where $\lambda$ is the London penetration depth,
and has periodic boundary conditions in the $x$ and $y$ directions.
Note that according to standard simulation practice,
for a square pinning lattice,
the sample is of equal size in the $x$ and $y$ directions,
but for a triangular pinning lattice,
the length of the sample along the $y$ direction is reduced
slightly in order to accommodate a triangular lattice without any defects.
We fix the pinning density to $n_p=N_p/L^2=0.5$
and the pinning radius to $r_{p} = 0.35$.
We have previously used these parameters to investigate
commensuration effects for bulk driven vortices, where
peaks in the critical depinning force
appear at commensurate fields \cite{Reichhardt98a,Reichhardt01a,Reichhardt07a,Reichhardt09}. 

For the colloidal system, we use similar overdamped dynamics
but replace the Bessel function interaction by 
a screened Coulomb or Yukawa potential
of the form $U(r_{ij}) \propto Q^2 \exp(-\kappa r_{ij})/r_{ij}$,
where $Q$ is the charge on an individual colloidal particle
\cite{Reichhardt02,Brunner02}. Here, the strength of
the colloid-colloid interaction can be varied
by changing the charge $Q$, such as by modifying the
ion concentration of the solution.  

\section{Results}

\begin{figure}
\includegraphics[width=3.5in]{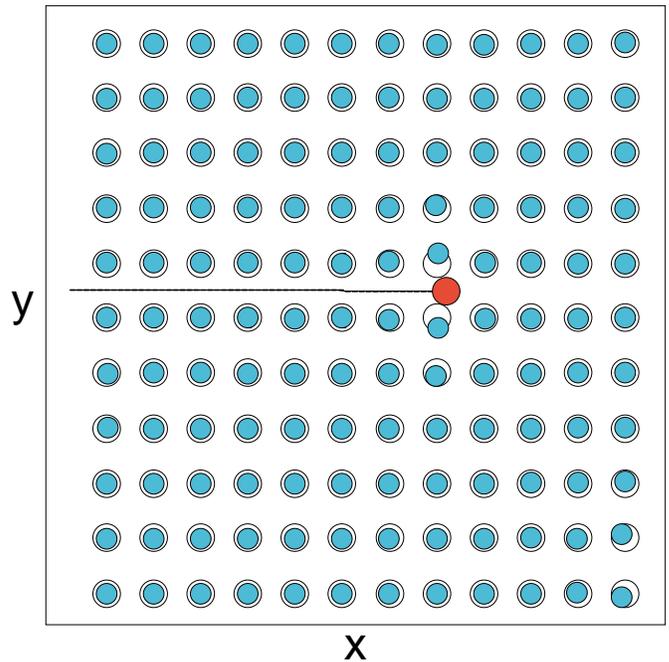}
\caption{ 
Image of a subsection of the superconducting
vortex system showing a square pinning array (open circles),
the bulk vortices (blue circles), 
the driven vortex (red circle), and the vortex trajectories (lines)
in a sample with $f=B/B_{\phi} = 1.0$ or a 1:1 matching of bulk vortices to pinning
sites.
Here  $F_{D} = 1.0$ and $F_{p} = 0.25$.
The size of the pinning sites has been adjusted for clarity.
	}
\label{fig:1}
\end{figure}

We first consider a superconducting system with a single driven vortex.
In Fig.~\ref{fig:1} we plot a subsection of the sample showing
the positions of the bulk vortices and pinning sites along with the trajectory
of the driven vortex
at $f=B/B_{\phi} = 1.0$ and $F_{p} = 0.25$.
Here the bulk vortices
form an ordered commensurate state
in which
each pinning site captures one vortex and 
the driven probe particle moves along a 1D channel
in the interstitial region.

\begin{figure}
\includegraphics[width=3.5in]{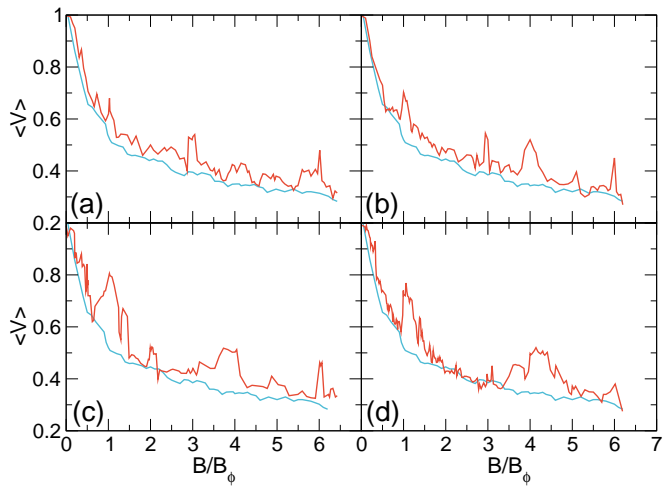}
\caption{
(a) The average velocity $\langle V\rangle$  of the driven particle
  versus $B/B_{\phi}$  at $F_{D} = 1.0$.
The blue line is the same in all panels and is from a sample
with
$F_{p} = 0.0$ or no pinning, which shows
a monotonic decrease in $\langle V\rangle$ with increasing $B/B_{\phi}$.
The red lines are for finite pinning.
(a) A square pinning array with $F_{p} = 0.03$.
(b)
A square pinning array with $F_{p} = 0.25$,
where there are velocity peaks at $B/B_{\phi} = 1/2$, 1.0, 2.0, 4.0, and
$6.0$, along with weaker peaks at $B/B_{\phi} = 2.0$ and $5.0$.
(c) A triangular pinning array with $F_p=0.03$,
where there are velocity peaks at $B/B_{\phi} = 1.0$, 3.0, 4.0, and $6.0$,
as well as a missing peak at $B/B_{\phi} = 2.0$.
(d) A triangular pinning array with $F_p=0.25$.
In general, the samples containing
finite pinning also
have a larger average $\langle V\rangle$ compared to samples without pinning.  
}
\label{fig:2}
\end{figure}

In Fig.~\ref{fig:2}(a) we plot the average velocity of the driven particle
$\langle V\rangle$ at an applied drive of 
$F_{D} = 1.0$ as a function of $B/B_{\phi}$
for a square pinning array with $F_{p} = 0.0$ (no pinning) and
$F_{p} = 0.03$. When there is no pinning, $\langle V\rangle$ decreases monotonically
approximately as $\langle V\rangle \propto B^{1/2}$
from $\langle V\rangle = 1.0$ at $B/B_{\phi} = 0.0$ to
$\langle V\rangle = 0.3$ at $B/B_{\phi} = 6.0$.
This behavior arises due to the increased number and strength
of interactions between the driven
vortex and the background vortices as the
magnetic field, and thus the vortex density, increases.
In the presence of a square pinning array with
$F_{p} = 0.03$, small peaks appear in $\langle V\rangle$ at
$B/B_{\phi} = 1.0$, 3.0, 4.0, and $6.0$. 
At these fillings, the system forms an ordered lattice.
From previous studies of vortices in a square periodic pinning array,
it is known that ordered lattices appear only for
certain matching fields and not for all matching fields,
while the particle ordering
also depends on the pinning strength \cite{Reichhardt98a}. 
Specifically, for a square pinning array,  
a square vortex lattice is stabilized at $B/B_{\phi}  = 1.0$, 2.0, and 5.0,
a triangular vortex lattice appears at $B/B_{\phi} = 4.0$ and $6.0$,
and peaks in the depinning force occur at all of these fields
under bulk driving \cite{Reichhardt98a}.
For the relatively weak square pinning lattice with
$F_{p} = 0.03$ in Fig.~\ref{fig:2}(a),
we find
a square vortex lattice at $B/B_{\phi} = 2.0$ but
a triangular vortex lattice at $B/B_{\phi} = 3.0$ and $B_{\phi} = 6.0$.
In Fig.~\ref{fig:2}(b) we plot $\langle V\rangle$ versus $B/B_{\phi}$ for
a square pinning array with a stronger  
pinning force of $F_{p} = 0.25$, where the orderings are more similar to those found
in previous studies \cite{Reichhardt98a}, with a
vortex lattice that is square at $B/B_{\phi} = 1.0$, 2.0, and $5.0$ and
triangular
at $B/B_{\phi} = 4.0$ and $6.0$.
At $B/B_{\phi} = 3.0$, the vortex lattice is neither square nor triangular, but
has dimer ordering
\cite{Reichhardt98a}, while
at $B/B_{\phi} = 5.0$, the vortex lattice is mostly square but contains
numerous grain boundaries.
Fields with square or triangular vortex lattices are
accompanied by a peak in the vortex velocity,
and there is a missing velocity peak at $B/B_{\phi}=3.0$.
In previous work with bulk driven systems, the formation of
a square vortex lattice under the influence of strong pinning is correlated
with the appearance of
peaks in the {\it critical depinning force} and a
{\it reduction} in the velocity of flowing vortices.
There was also a missing peak in the depinning force at $B/B_{\phi} = 3.0$.  
The previous work combined with our new results indicates that
the formation of ordered vortex lattices
at commensurate fields produces 
enhanced pinning (indicated by a peak in the critical depinning and a dip in
the vortex velocity) for bulk driven systems
but reduced pinning (indicated by a peak in the vortex velocity)
for individually driven vortices.
We note
that there is also a submatching velocity peak at $B/B_{\phi} = 1/2$ 
in Fig.~\ref{fig:2}(b).

In Fig.~\ref{fig:2}(c) we plot $\langle V\rangle$
versus $B/B_{\phi}$ for the pin-free sample and for a sample containing a
weak triangular pinning lattice 
with $F_{p} = 0.03$.
In the presence of pinning,
peaks in the velocity appear
at $B/B_{\phi}  = 1.0$, 3.0, and $6.0$
that are more prominent than the peaks observed for a square pinning lattice.
In previous work on bulk driven vortices, commensurate effects
produce enhanced pinning (rather than enhanced velocity)
at $B/B_{\phi} = 1.0$, $3.0$, and $4.0$,
but not at $B/B_{\phi} = 3.0$, since for the latter filling
the vortices form a partially ordered honeycomb lattice rather than a triangular lattice. 
For the individually driven vortex, Fig.~\ref{fig:2} indicates
that the net velocity is generally lower in the absence of pinning and higher
in the presence of pinning, which is the opposite of the behavior generally found
for bulk driving.
Additionally, the peak velocity is shifted slightly above the matching field rather
than occurring exactly at the matching field.
In Fig.~\ref{fig:2}(d), we plot $\langle V\rangle$ versus $B/B_{\phi}$
for a stronger triangular pinning array with $F_{p} = 0.25$.
Here there is a broad
velocity enhancement peak over the range $3.0 < B/B_{\phi} < 4.5$, 
where
the system forms a triangular lattice with varied orientations. 

The overall behavior in Fig.~\ref{fig:2} is the opposite of what is found
for bulk driving, 
where at commensurate fillings the driven particles are less coupled to
the substrate and move more rapidly.
The drag on an individually driven particle arises due to
interactions with the surrounding background particles,
and for incommensurate fillings, these particles
are more disordered and less coupled to the substrate, permitting them to
couple more 
effectively to the driven particle and generate a larger drag effect.
When pinning is present, the background particles couple to the substrate
and cannot be entrained as easily by the driven particle, which reduces
the drag.
Under commensurate conditions, the surrounding particles are very strongly
coupled to the substrate and poorly coupled to the driven particle.
As a result, the driven particle
is more mobile at the matching fields,
which we call
an anti-commensuration effect. 
The peaks in the velocity fall at fields slightly above matching because
just below matching,
the background particles form a commensurate lattice
with vacancies
that act as effective trapping sites for the driven particle.
In contrast, slightly above commensuration, there are some interstitials in the
commensurate configuration which are unable to trap the driven particle.

\begin{figure}
\includegraphics[width=3.5in]{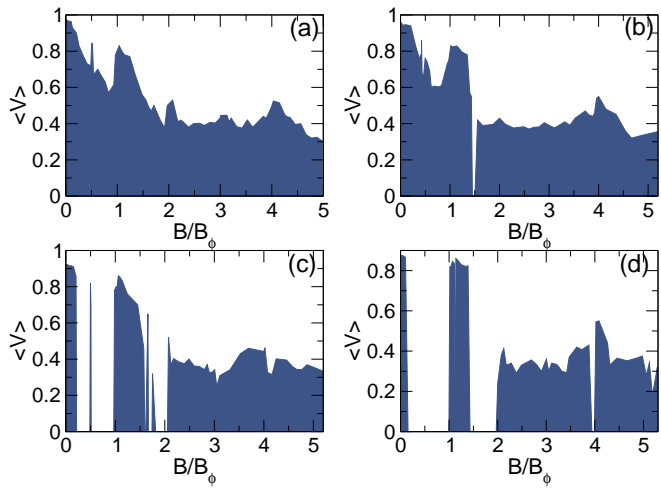}
\caption{
$\langle V\rangle$ versus $B/B_{\phi}$ for the system in
Fig.~\ref{fig:2}(a) with a square pinning array at $F_D=1.0$ and
(a) $F_{p} = 0.375$, (b) $F_{p} = 0.5$, (c) $F_{p} = 0.625$, and (d) $ F_{p} = 0.75$.
Here the commensurate conditions correspond to fields at which
there is a peak in the velocity. For larger $F_{p}$, 
there are windows of field over which the system becomes pinned with $\langle V\rangle=0$. 
}
\label{fig:3}
\end{figure}

In Fig.~\ref{fig:3} we plot $\langle V\rangle$ versus $F_{D}$ for the system
in Fig.~\ref{fig:2}(a,b) with a
square pinning lattice at different pinning strengths.
At $F_{p} = 0.375$ in Fig.~\ref{fig:3}(a), strong velocity peaks appear at
$B/B_{\phi} = 1/2$, 1.0, 2.0, and $4.0$,
while for $F_{p} = 0.5$ in Fig.~\ref{fig:3}(b), the peak locations remain robust but
several regions have opened in which the driven particle is pinned with
a velocity of $\langle V\rangle=0$.
One pinned region appears just above
$B/B_{\phi} = 1/2$, while another is found near $B/B_{\phi} = 1.5$.
For $F_{p}= 0.625$ in Fig.~\ref{fig:3}(c), there are a growing number of
pinned regions, while peaks in $\langle V\rangle$ appear at
$B/B_{\phi} = 1/2$, $1.0$, and just above $2.0$.
For
$F_{p} = 0.75$ in Fig.~\ref{fig:3}(d),
the pinned regions have expanded
while velocity peaks persist at $B/B_{\phi}  = 1.2$ and just above $B/B_{\phi}=2.0$ and
$B/B_{\phi}=4.0$.
These results indicate that
strong pinning effectiveness appears in the incommensurate phase,
which is the opposite of the behavior observed in a bulk driven system.
Since $F_D>F_p$ for all of the samples in Fig.~\ref{fig:3},
the effective drag or pinning of the driven vortex is produced by a
combination of trapping directly at pinning sites and interactions with
background pinned vortices.

\begin{figure}
\includegraphics[width=3.5in]{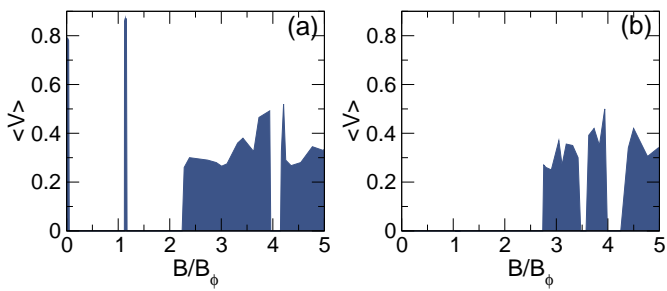}
\caption{ 
$\langle V\rangle$ versus $B/B_{\phi}$ for the system in
Fig.~\ref{fig:2}(a) with a square pinning array at $F_D=1.0$ at
(a) $F_{p} = 0.875$ and (b) $F_{p} = 1.0$, where the widths of the
pinned regions increase with increasing $F_p$.
}
\label{fig:4}
\end{figure}

In Fig.~\ref{fig:4}(a,b) we show $\langle V\rangle$ versus $B/B_\phi$ for
samples with a
square pinning array at higher pinning strengths of
$F_{p}= 0.875$ and $1.0$.
At $F_{p} = 0.875$ in Fig.~\ref{fig:4}(a),
there is a small region close to $B/B_{\phi} = 0.0$ where the
driven vortex only interacts with the pinning sites and remains unpinned
since $F_D>F_p$.
In this regime, $\langle V\rangle  = 0.79 < F_D$ since the driven particle
is still slowed by the interactions with the pinning sites.
Just above $B/B_{\phi} = 1.0$, we find another pinned region
which extends up to $B/B_{\phi} = 2.5$.
For a higher $F_p=1.0$ in Fig.~\ref{fig:4}(b), individual pinning sites can
trap the driven vortex since we now have $F_p=F_D$,
and a large pinned regime extends from $0 \leq B/B_{\phi} < 2.75$.

\begin{figure}
\includegraphics[width=3.5in]{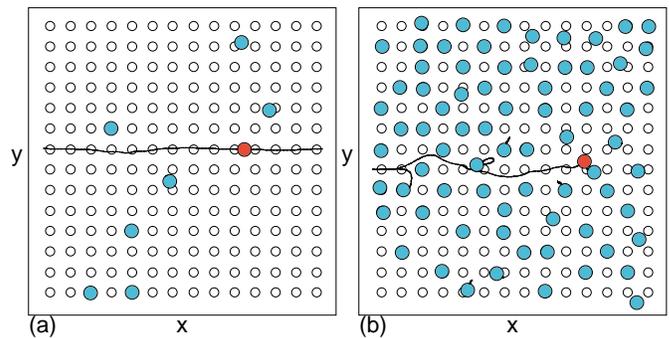}
\caption{Image of a subsection of the superconducting vortex system with
  a square pinning array (open circles), bulk vortices (blue circles), the driven
  vortex (red circle), and the vortex trajectories (lines)
  for the system in
 Fig.~\ref{fig:3}(a) with $F_{p} = 0.375$ and $F_D=1.0$. 
The size of the pinning sites has been adjusted for clarity.
 (a) $B/B_{\phi} = 0.05$ where the driven particle moves along a pinning row
 and has little interaction with the other vortices.
 (b) $B/B_{\phi}  = 0.414$, an incommensurate regime,
 where the flow is much more disordered with multiple collisions occurring between
 the driven vortex and the background vortices. 
	}
\label{fig:5}
\end{figure}

\begin{figure}
\includegraphics[width=3.5in]{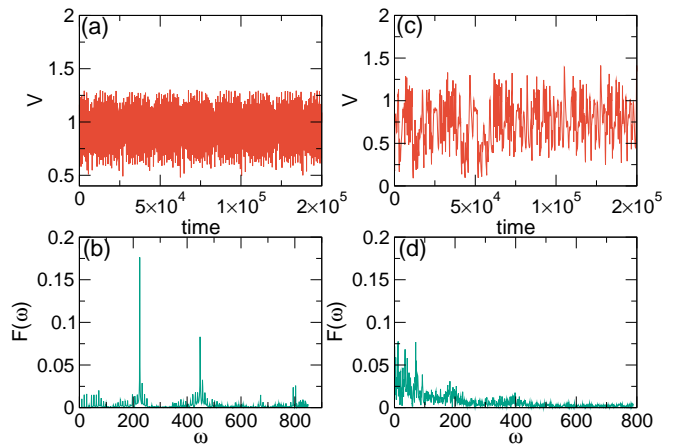}
\caption{
(a, c) Time series of the driven particle velocity $V$
  and (b, d) the corresponding
  Fourier transform $F(\omega)$ for the system in
Fig.~\ref{fig:5} with a square pinning array at $F_p=0.375$ and $F_D=1.0$. 
(a, b) At $B/B_{\phi} = 0.05$,  the motion is periodic,
producing a series of peaks in $S(\omega)$.
(c, d) At $B/B_{\phi} = 0.414$, the velocity signal is more disordered  and the
corresponding $F(\omega)$ has much broader noise features.
	}
\label{fig:6}
\end{figure}

To characterize the dynamics of the driven particle, we examine
detailed velocity time series as well as images of the sample.
In Fig.~\ref{fig:5}(a) we illustrate
the system in Fig.~\ref{fig:3}(a) at 
$F_{p} = 0.375$ for a low field of $B/B_{\phi} = 0.05$,
where the driven particle motion is localized along
a single row of pinning, and there are only weak interactions with
the background vortices.
In Fig.~\ref{fig:6}(a) we plot the time series of the velocity
$V(t)$ for the sample
in Fig.~\ref{fig:5}(a).
The signal is periodic since the driven vortex moves directly over the
pinning sites in a single row, while there are modulations of the signal at
longer time scales due to the occasional interactions with the background vortices.
Figure~\ref{fig:6}(b) shows the Fourier transform $F(\omega)$ of the time
series in Fig.~\ref{fig:6}(a),
\begin{equation}
	F(\omega) = \int V(t)e^{-i\omega t}dt \ .
\end{equation}
Here $F(\omega)$ exhibits
a strong periodic or narrow band noise feature due to the
periodic motion of the driven vortex.
Figure~\ref{fig:5}(b) shows the vortex trajectories at
$B/B_{\phi} = 0.414$ in the incommensurate region, where the motion
of the driven particle is much more random due to strong interactions with
numerous background vortices, which produce considerable random winding
of the trajectory
in the direction transverse to the drive.
The driven vortex follows a different trajectory each time it passes through the
periodic boundary conditions, suggesting that the flow has a chaotic character.
The corresponding velocity time series in Fig.~\ref{fig:6}(c) contains strong
random  fluctuations, while the Fourier transform $F(\omega$)
in Fig.~\ref{fig:6}(d) is broad and has lost the sharp peaks observed at
$B/B_{\phi} = 0.05$.  
When $B/B_{\phi} = 0.5$, the background vortices adopt an ordered checkerboard
ordering \cite{Grigorenko03,Reichhardt01a}, and the driven particle moves along
an ordered 1D path without collisions, similar to the trajectory shown in Fig.~\ref{fig:1}.

\begin{figure}
\includegraphics[width=3.5in]{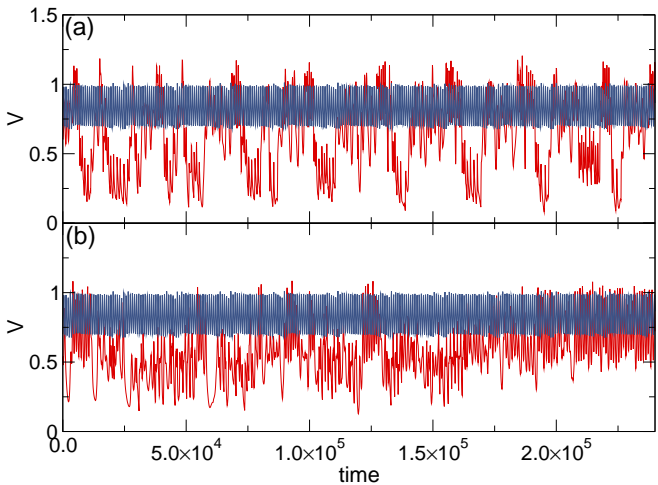}
\caption{
  (a) Time series of the velocity
$V$  of the driven particle for the system from Fig.~\ref{fig:1} with
$F_p=0.375$ and $F_D=1.0$ at 
  $B/B_{\phi} = 1.0$ (blue)
and $B/B_{\phi} = 0.828$ (red).
(b) The same for
  $B/B_{\phi} = 1.0$ (blue) and
  $B/B_{\phi} = 1.55$ (red).
  In each case, the velocity at the commensurate filling of $B/B_{\phi}=1$ is
  periodic and has a higher average value than the velocity at the
  incommensurate fillings.
}
\label{fig:7}
\end{figure}

To illustrate more clearly the change in the velocity fluctuations of the driven
particle depending upon whether the filling is commensurate or not,
in Fig.~\ref{fig:7}(a) we plot a time series of the velocity $V$ of the driven
particle for the system from Fig.~\ref{fig:1} with $F_p=0.375$ and $F_D=1.0$
at a commensurate state of $B/B_{\phi} = 1.0$
and an incommensurate state of $B/B_{\phi} = 0.828$.
For the incommensurate field, the average velocity is lower  and
the signal is more disordered,
while for $B/B_{\phi} = 1.0$, the signal is periodic and has a higher 
average
value.
At $B/B_{\phi} = 1.0$, the driven particle moves in a 1D path as illustrated in
Fig.~\ref{fig:1},
and the oscillatory part of the velocity signal arises
from the periodic interactions 
of the driven particle with the
vortices at the pinning sites.
Even though there are more background vortices with which the driven
particle could interact at $B/B_{\phi}=1.0$ than at
$B/B_{\phi} = 0.828$,
the velocity is higher for $B/B_{\phi}=1.0$ since the ordered nature of the commensurate
flow reduces the number of effective collisions between the driven particle and
the background vortices.
In Fig.~\ref{fig:7}(b) we compare the velocity time series in the same sample
at $B/B_{\phi}=1.0$ with that of the higher incommensurate field $B/B_{\phi}=1.55$.
Here, the velocity is again reduced in magnitude and
more chaotic in nature at the incommensurate filling
than at the $1:1$ matching field.  

\begin{figure}
\includegraphics[width=3.5in]{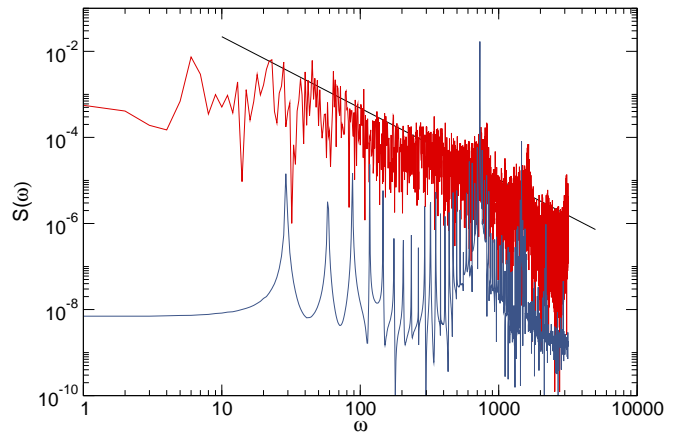}
\caption{(a)
The power spectra $S(\omega)$ 
for	the samples in Fig.~\ref{fig:7}(a)
with $F_p=0.375$ and $F_D=1.0$
at the commensurate field
$B/B_{\phi}=1.0$ (blue), where the noise is strictly narrow band,
and at the incommensurate field $B/B_{\phi}=0.828$,
where the noise is broad.
The solid line is a fit to $S(\omega) \propto \omega^{-1.75}$.
}
\label{fig:8}
\end{figure}

In Fig.~\ref{fig:8} we show the power spectrum $S(\omega)$,
which is the absolute square of $F(\omega)$, for
the samples in Fig.~\ref{fig:7}(a).
At the commensurate field of
$B/B_{\phi}  = 1.0$,
$S(\omega)$ has
a strict narrow band character,
while at the incommensurate field of $B/B_{\phi}  = 0.828$,
the velocity noise is broad band, as indicated by
the solid line which is a fit to $S(\omega) \propto \omega^{-1.75}$.
There are still some weaker periodic peaks
for $B/B_{\phi} = 0.828$ since there is still
a periodic component of the velocity induced by the substrate, and the
positions of these peaks match the positions of the peaks found for
$B/B_{\phi} = 1.0$; however, the overall noise power is much larger for the
incommensurate system, indicating that the driven particle velocity fluctuations
extend out to much longer time scales.
The presence of broad band noise or $\omega^{-\alpha}$ noise  
in driven systems with random disorder is associated with plastic or disordered flow, while
narrow band noise is associated with ordered flow \cite{Reichhardt17}.
The velocity noise power spectrum for the incommensurate system
with $B/B_{\phi} = 1.55$ has broad band features similar to those shown for
$B/B_{\phi} = 0.828$. 
Our results indicate that at the incommensurate fields, the
flow has plastic characteristics since the driven particle generates
position exchanges among the background vortices.
Due to the periodic boundary conditions,
the driven particle moves repeatedly through the sample; however, 
during each passage
the vortex positions are rearranged by plastic events,
so the landscape experienced by the driven particle
changes over time, producing
the large low frequency velocity noise.
In contrast, under commensurate conditions
the background particles are fixed
and the driven particle experiences the same background during each
pass through the sample.
It is also possible for strictly
periodic flow to occur for fillings slightly away from commensuration,
$B/B_{\phi}=1.0 \pm \epsilon$, when
a small number of vacancies or interstitials are present in the background lattice.
As long as $\epsilon$ is small enough that the vacancies or interstitials are well
spaced and immobile,
the driven particle experiences a periodic potential from the background lattice at
one frequency, along with
lower frequency periodic  perturbations from 
the interstitials or vacancies.
When $\epsilon$ becomes too large, interactions among the interstitials or
vacancies begin to occur and plastic deformations of the background lattice
become possible, causing the system to transition to a
broad band velocity noise regime of the type shown
in Fig.~\ref{fig:8}.

\begin{figure}
\includegraphics[width=3.5in]{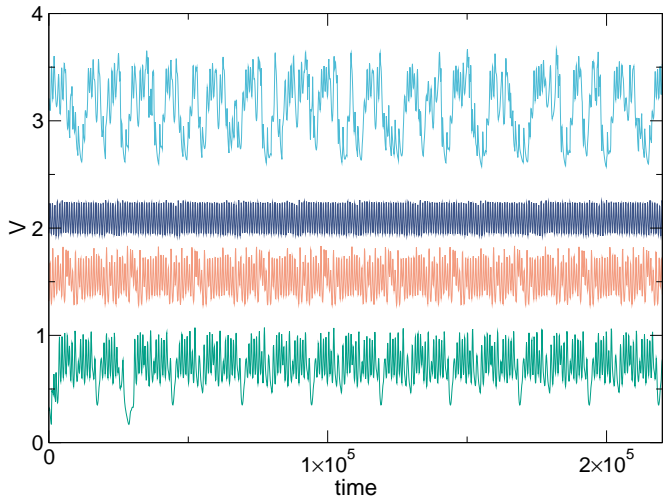}
\caption{(a)
The time series of the driven particle velocity $V$ for the system in
Fig.~\ref{fig:7}(a) with $F_p=0.375$ and $F_D=1.0$
at $B/B_{\phi} = 0.9328$ (light blue), $1.0$ (dark blue),
$1.24$ (orange),  and $1.29$ (green).
The curves have been shifted vertically for clarity.
The most ordered motion occurs for $B/B_{\phi} = 1.0$.  
        }
\label{fig:9}
\end{figure}

In Fig.~\ref{fig:9} we plot the
velocity time series for filling fractions at and on either side of
the $B/B_{\phi} = 1.0$ matching field,
where the velocities have been
shifted vertically for clarity. 
The velocity fluctuations below matching for $B/B_{\phi} = 0.9328$ are much
stronger than those found
above matching for $B/B_{\phi}=1.24$ and $B/B_{\phi}=1.29$.
Here, the vacancy sites which form below matching can
act as trapping sites for the driven particle, permitting strong perturbations
of the motion to occur.
In contrast,
for $B/B_{\phi} = 1.24$ the average velocity is relatively high and the fluctuations
are reduced in magnitude, since these fluctuations are produced by the much
weaker interactions of the driven particle with 
interstitial vortices.
For $B/B_{\phi} = 1.29$, the density of interstitial vortices has become large enough
that the interstitials begin to interact with each other, permitting plastic motion
to occur and resulting in stronger
periodic  velocity drops of the driven particle. 

\begin{figure}
\includegraphics[width=3.5in]{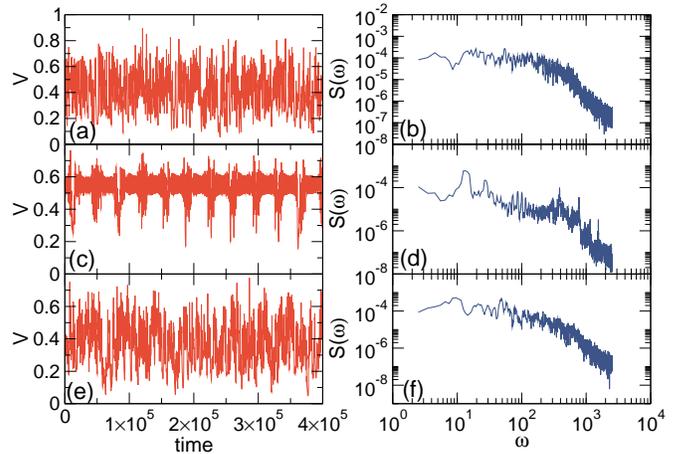}
\caption{(a, c, e) The velocity time series $V(t)$ and
(b, d, f)  the corresponding power spectra $S(\omega)$
for the system in Fig.~\ref{fig:3}(a) with $F_{p} = 0.375$ and $F_D=1.0$.
(a,b) $B/B_{\phi} = 3.0$,
where the system is disordered. 
(c,d)
$B/B_{\phi} = 4.03$, where the system is ordered.
(e,f)
$B/B_{\phi} = 4.4576$, where the vortex
positions are  more disordered and the velocity power spectrum has a
broad band character.
	}
\label{fig:10}
\end{figure}

In general, we find that for commensurate states associated with peaks in the
average velocity $\langle V\rangle$,
the flow of the driven particle is ordered
and exhibits a periodic velocity signature,
while at incommensurate fillings, the flow is plastic
and has a broad band noise signal.
In Fig.~\ref{fig:10}(a,b) we plot the velocity time series $V(t)$ and
power spectrum $S(\omega)$
for the system in Fig.~\ref{fig:3}(a) with $F_{p} = 0.375$ and $F_D=1.0$ at
$B/B_{\phi} = 3.0$.
For this filling, there is no peak in $\langle V\rangle$ and
the background vortices do not form an ordered square or triangular lattice but
instead adopt a disordered structure.
As the probe particle passes through the sample,
it generates strong plastic deformations of the background vortices, and
the velocity noise has a broad band character.
For $B/B_{\phi}=4.03$ in Fig.~\ref{fig:10}(c,d),
slightly above the fourth matching field where a peak
in $\langle V\rangle$ appears in Fig.~\ref{fig:3}(a),
the background vortices
form a triangular lattice that contains a few grain boundaries.
Here, the driven particle spends most of its time traversing
ordered portions of the lattice, and exhibits a dip in $V$ each time it
crosses one of the grain boundaries.
The power spectrum indicates that two separate periodicities are
combined in the velocity signal.
The higher frequency peaks are associated with the motion of the driven
particle through the periodic commensurate vortex lattice,
while the lower frequency narrow band
noise is produced by the periodic motion of the probe particle
over the grain boundaries.
A given grain boundary
can move gradually over time since it can be entrained briefly by the
probe particle, producing a drop in the probe
particle velocity.
At $B/B_{\phi}=4.4576$ in Fig.~\ref{fig:10}(e,f),
the vortex positions are  more disordered
and the velocity time series has chaotic features,
while the power spectrum has a broad band noise shape.

\begin{figure}
\includegraphics[width=3.5in]{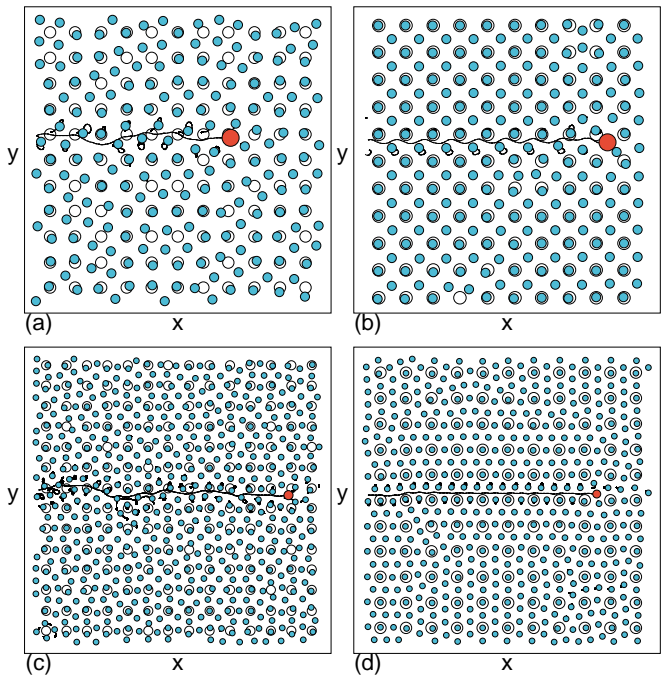}
\caption{
Image of a subsection of the superconducting vortex system with a square
pinning array (open circles), bulk vortices (blue circles), the driven vortex (red circle),
and the vortex trajectories for
a system with 
$F_p=0.375$ and $F_D=1.0$.
The size of the pinning sites has been adjusted for clarity.
(a) Disordered flow at $B/B_{\phi} = 1.55$.
(b) Ordered flow at $B/B_{\phi} = 2.0$.
(c) Disordered flow at $B/B_{\phi} = 3.0$.
(d) Ordered flow at $B/B_{\phi} = 4.03$.
}
\label{fig:11}
\end{figure}

In Fig.~\ref{fig:11}(a) we show the bulk vortices, pinning sites, driven vortex,
and vortex trajectories for a portion of a system with $F_p=0.375$ and $F_D=1.0$
at $B/B_{\phi} = 1.55$.
The background vortices are disordered and the trajectory of the probe
particle is also disordered,
giving the strongly fluctuating velocity signature shown
in Fig.~\ref{fig:7}(b) and producing broad band velocity noise.
At $B/B_{\phi} = 2.0$, where there is a peak in $\langle V\rangle$
and the background vortices form an ordered square lattice,
Fig.~\ref{fig:11}(b) indicates that
the driven particle moves through the interstitial region
along a sinuous path modulated by the interstitial vortices, resulting in a
narrow band noise signature.
In Fig.~\ref{fig:11}(c)
at $B/B_{\phi} = 3.0$,
the background vortices 
are not crystalline, consistent with the absence of a peak in
$\langle V\rangle$ at this filling.
The driven vortex follows a disordered path and exhibits broad band
noise.
Figure~\ref{fig:11}(d) shows that for
$B/B_{\phi} = 4.03$, 
the background vortices form a polycrystalline lattice and the
flow is mostly ordered, giving the narrow band noise illustrated 
in Fig.~\ref{fig:10}(c,d).

\begin{figure}
\includegraphics[width=3.5in]{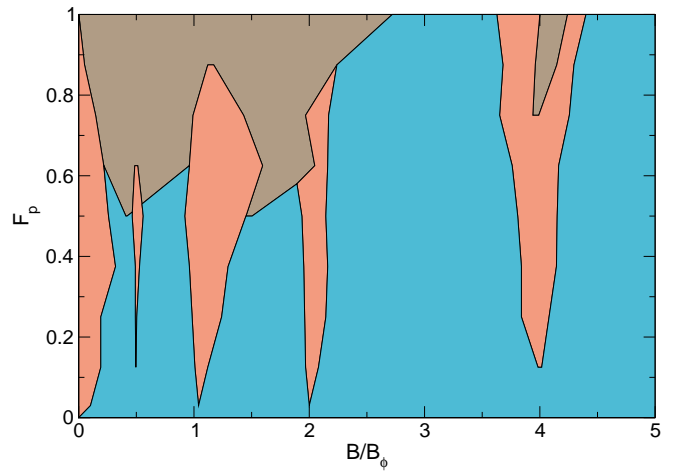}
\caption{(a)
Dynamic phase diagram as a function of pinning strength $F_{p}$
versus filling $B/B_{\phi}$ highlighting the pinned phase (brown),
the viscous flow or broad band noise phase (blue),
and the periodic flow or narrow band noise phase (orange)
for the system from Figs.~\ref{fig:3} and \ref{fig:4} with
$F_D=1.0$.
Only the largest pinned regions are marked; smaller pinned regimes are not shown.
	}
\label{fig:12}
\end{figure}

Based on the features in $\langle V\rangle$ and $S(\omega)$,
we construct a dynamic phase diagram
as a function of pinning strength $F_p$ versus $B/B_{\phi}$ with $F_D=1.0$
in Fig.~\ref{fig:12}, where we
highlight the pinned, viscous flow or broad band noise,
and narrow band noise regimes.
Only the largest pinned phases are marked; there are several smaller pinned
phases which are not shown on the figure.
For finite $F_{p}$, at low  fillings
the system is always in a periodic motion state since the driven
particle moves along the pinning rows as shown in Fig.~\ref{fig:5}(a).
Similar periodic oscillations arise 
around $B/B_{\phi} = 1/2$,  1.0, 2.0, and $4.0$. 
The pinned regions reach their greatest extent for incommensurate fillings,
consistent with our observation that the effectiveness of the pinning is
generally reduced at commensurate fillings.
For low fillings, the driven particle does not become pinned until
$F_D \geq F_{p}$.
For $B/B_{\phi} > 3.0$, there are windows of $B/B_{\phi}$ over which
the driven particle never becomes pinned,
since all of the pinning sites are  occupied
and the driven particle can only be pinned through caging by interactions
with neighboring vortices.
At lower pinning strengths, the flow is not periodic even at the matching 
fields since as the driven particle moves it can
interact strongly enough with the background vortices
to generate plastic flow events. 

\begin{figure}
\includegraphics[width=3.5in]{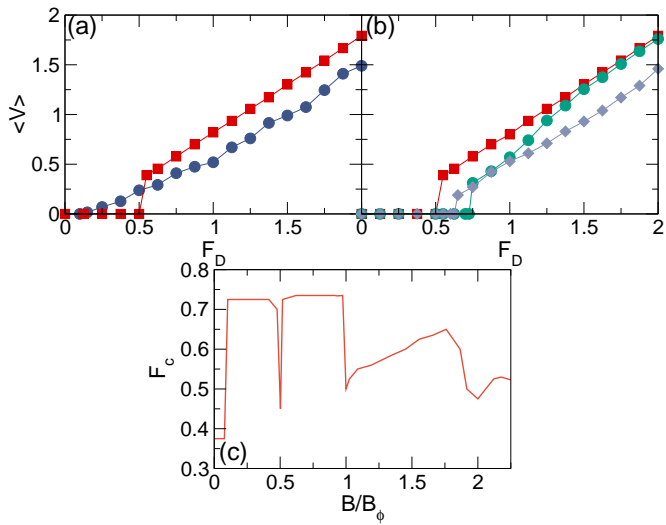}
\caption{(a)
$\langle V\rangle$ versus driving force $F_{D}$
for the system in Fig.~\ref{fig:3} at
$B/B_{\phi} = 1.0$
with $F_{p}  = 0.0$ (blue circles) and $F_{p} = 0.375$ (red squares). 
(b) $\langle V\rangle$ versus $F_{D}$ for the same system
at $F_{p} = 0.375$
for $B/B_{\phi} = 0.9328$ (green circles),
$1.0$ (red squares), and $1.55$ (light blue diamonds).
(c) The critical depinning force $F_{c}$ versus $B/B_{\phi}$ for the same system. 
	}
\label{fig:13}
\end{figure}

We next consider the drive dependence and pinning
transition for the driven particle.
In Fig.~\ref{fig:13}(a) we plot the
velocity-force curve $\langle V\rangle$ versus $F_{D}$
of the driven vortex for the system in Fig.~\ref{fig:3}
at $B/B_{\phi} = 1.0$
for $F_p=0.0$ and $F_p=0.375$.
In the sample with no pinning,
there is still a finite depinning transition 
just above $F_{D} = 0.1$ produced by the caging of the driven vortex due to
its interactions with the surrounding vortex lattice.
When $F_{p} = 0.375$, the depinning transition
occurs at $F_{p} = 0.5$, about five times higher than in the sample
without pinning, and is much sharper.
In the pin-free sample,
the depinning transition is plastic and depinning occurs when
vortices in the surrounding lattice become able to exchange places
with each other.
For the finite pinning sample,
the depinning is elastic and the driven particle moves
without creating any plastic distortions in the surrounding lattice.
This motion can be
viewed as
a single particle traveling over a 
fixed square 2D substrate.
Once the particle is moving, $\langle V\rangle$
is higher for the sample containing pinning than for the sample without pinning.
The pinning prevents the surrounding vortices from being dragged by the
driven particle or from exchanging places with each other in plastic events,
both of which are processes which increase the drag on the driven particle.
This is consistent with the results in Fig.~\ref{fig:2} where
$\langle V\rangle$ is lower for a sample
with finite pinning 
than for a sample with no pinning.

In Fig.~\ref{fig:13}(b) we plot
$\langle V\rangle$ versus $F_{D}$ for the system with $F_{p} = 0.375$
at $B/B_{\phi} = 0.9328$,
$1.0$,
and $1.55$.
The depinning threshold is lowest when $B/B_{\phi} = 1.0$,
and in the flowing phase, $\langle V\rangle$ is smaller at incommensurate fillings
than at the commensurate filling.
Figure~\ref{fig:13}(c) shows the critical depinning force
$F_{c}$ versus $B/B_{\phi}$ obtained  from a series of 
velocity-force curves for the system in Fig.~\ref{fig:13}(b).
For $B/B_{\phi} < 0.075$, we find $F_{c} = F_{p}$ since the
depinning occurs in the single particle limit and is controlled
only by direct depinning from the pinning sites, with vortex-vortex interactions
playing no role.
For $0.075 \leq B/B_{\phi} < 1/2$,
the depinning threshold reaches its largest value of $F_{c} = 0.725$,
produced by a combination of direct pinning effects and additional interstitial
pinning due to interactions with the neighboring vortices.
At $B/B_{\phi} = 1/2$, where  the system forms a checkerboard  
ordered state, the driven particle moves
along a 1D path between the pinning sites and the 
only pinning arises from interactions with other vortices.
Here the driven particle does not move directly into a pinned
vortex; instead, it flows around the side of the pinned vortex where it experiences
a smaller repulsive force.
For $1/2 < B/B_{\phi} < 1.0$, the  system is disordered and the
driven particle experiences
pinning from both the pinning sites and the pinned vortices.
The depinning force drops again
at $B/B_{\phi} = 1.0$
when the driven particle follows a 1D path
between the pinning sites, as illustrated
in Fig.~\ref{fig:1}.
For $B/B_{\phi} > 1.0$, most of the pinning sites are occupied so
$F_{c}$ is generally lower.
There is
another drop in $F_c$ at $B/B_{\phi}  = 2.0$ where
the driven particle traverses a quasi-1D path between the pinning rows,
as shown in Fig.~\ref{fig:11}(b), while
for $B/B_{\phi} > 2.0$, $F_{c}$ increases again. 
This result indicates that the depinning threshold for the single driven vortex
is lowest at matching fields where there is an ordered configuration
of background vortices, which is opposite to the behavior found for bulk
driven vortices. 

\begin{figure}
\includegraphics[width=3.5in]{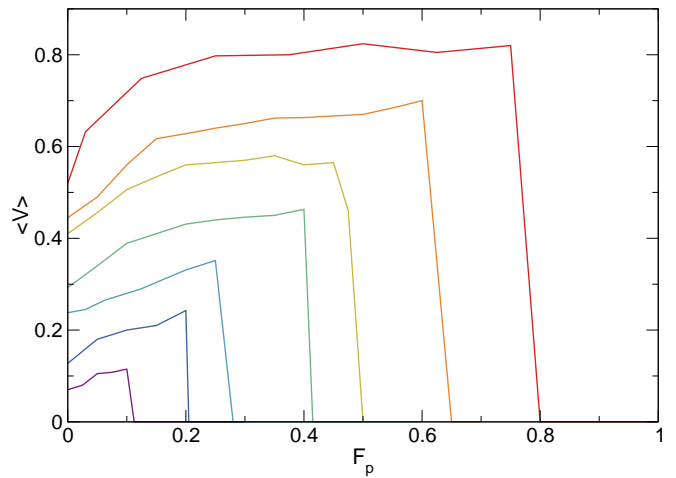}
\caption{
$\langle V\rangle$ versus $F_{p}$ at $B/B_{\phi} = 1.0$ for the system in
Fig.~\ref{fig:3} at
$F_{D} = 1.0$, 0.825, 0.75, 0.625, 0.5, 0.375, and $0.25$, from top to bottom,	
showing that for fixed drive, there is a regime in which
$\langle V\rangle$ 
increases with increasing $F_{p}$.
	}
\label{fig:14}
\end{figure}

For constant $F_D$
within the moving phase,
the velocity response increases with increasing $F_{p}$,
opposite to what is observed in most systems
with increasing pinning force \cite{Reichhardt17}.
In Fig.~\ref{fig:14} we plot
$\langle V\rangle$ versus $F_{p}$ at $B/B_{\phi} = 1.0$
for the system in Fig.~\ref{fig:3} at 
$F_{D} = 1.0$, 0.825, 0.75, 0.625, 0.5, 0.375, and $0.25$.
For a fixed driving force,
the velocity is low at $F_p=0$ and increases with
increasing $F_{p}$ before reaching a saturation and then dropping to zero
when $F_{p}$ becomes large enough.
For all the curves,
when $F_{p} < 0.06$ the system is in a
viscous or plastic flow regime since the pinning is no longer strong enough to hold
the background vortices
in place.  At higher 
$F_{p}$, the background vortices become pinned and the system enters
a periodic channel flow regime.  

\section{Triangular Pinning}

 \begin{figure}
\includegraphics[width=3.5in]{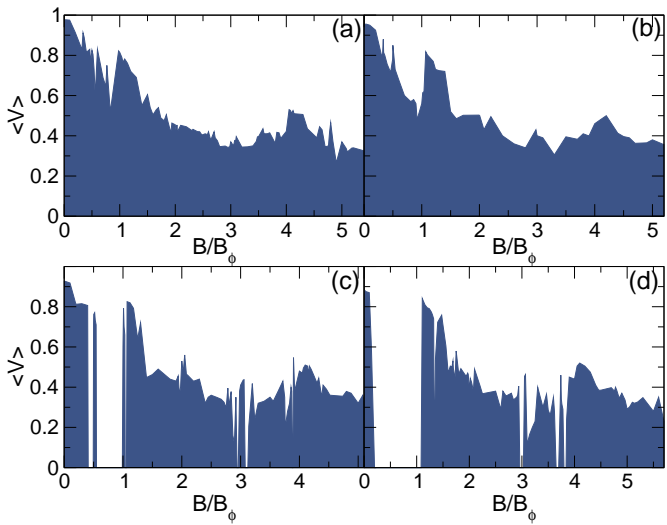}
\caption{(a)
$\langle V\rangle$ versus $B/B_{\phi}$ for a
 system with a triangular pinning array at $F_D=1.0$ and
$F_{p} =$ (a) 0.375, (b) 0.5, (c) 0.625, and (d) $0.75$.
 A series of peaks in
 $\langle V\rangle$ appear at matching fields
 for which the system forms an ordered lattice. 	
	 }
\label{fig:15}
 \end{figure}
 
\begin{figure}
\includegraphics[width=3.5in]{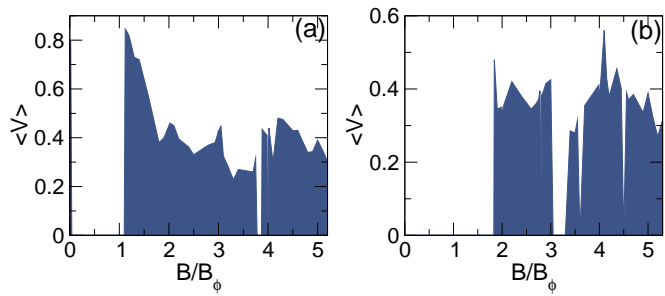}
\caption{(a)
$\langle V\rangle$ versus $B/B_{\phi}$ for
a system with a triangular pinning array
at $F_D=1.0$ and $F_{p}  =$ (a) 0.875 and (b) $1.0$.
	}
\label{fig:16}
\end{figure}

We next consider the
average velocity $\langle V\rangle$ of a single particle driven through a background of
vortices interacting with a
triangular pinning array
under constant drive.
In Fig.~\ref{fig:15} we plot $\langle V\rangle$ versus $B/B_{\phi}$ for
a system with a triangular pinning array at
$F_D=1.0$ and $F_{p} = 0.375$, 0.5, 0.625, and $0.75$. 
At $F_{p} = 0.375$ in Fig.~\ref{fig:15}(a),
there are large peaks in $\langle V\rangle$ at
$B/B_{\phi} = 1.0$ and $4.0$ along with smaller peaks at $B/B_{\phi} = 1/3$
and $3.0$.  There is no peak at $B/B_{\phi} = 1/2$ since for this filling the
background vortices form a
polycrystalline disordered state. 
For $F_{p} = 0.5$ in Fig.~\ref{fig:15}(b),
we observe the same trend,
but the peak at $B/B_{\phi} = 3.0$ 
is more prominent
since the system can achieve better ordering at this field
when the pinning is stronger.
Additionally,
a
peak begins to emerge at $B/B_{\phi} = 1/2$
due to the
formation of a stripe-like pattern.
From previous studies of vortex ordering on triangular arrays, it is known that
an ordered state can occur at
$B/B_{\phi}=1/3$ and a partially ordered
state can form at $B/B_{\phi}=1/2$ \cite{Reichhardt01a}.
At
$B/B_{\phi} = 3.0$ and $4.0$, the background vortices
form a triangular lattice,
but the orientation of this lattice at $B/B_{\phi} = 3.0$ involves the appearance of
a zig zag pattern in the interstitial region \cite{Reichhardt98a}
which makes it more difficult for the driven particle to pass through  the sample.
In contrast, at $B/B_{\phi} = 4.0$ the triangular lattice is oriented such that
there are 1D chains of interstitial vortices
along the $x$-direction \cite{Reichhardt98a},
making it possible for the driven particle to slide
easily alongside the chain without
weaving into the $y$ direction.
At $B/B_{\phi} = 5.0$, the background vortex positions are disordered and there
is no peak in $\langle V\rangle$.
Previous work showed
that a triangular lattice oriented similarly to the lattice observed
at $B/B_{\phi}=4.0$ appears
at $B/B_{\phi}=9.0$ but that there are two rows of interstitial vortices
instead of only one \cite{Reichhardt98},
so we expect that there would be another peak in $\langle V\rangle$
at this higher filling.
For $F_{p}  = 0.625$ in Fig.~\ref{fig:15}(c),
there are some regions
such as $1/2 < B/B_{\phi} < 1.0$.
where the system is pinned and $\langle V\rangle=0$.
A smaller velocity peak appears
at $B/B_{\phi} = 2.0$ where the background vortices attempt
to form an ordered honeycomb state \cite{Reichhardt98a},
while the peaks in $\langle V\rangle$ remain robust at $B/B_{\phi} = 3.0$ and $4.0$.
At $F_{p} = 0.75$ in Fig.~\ref{fig:15}(d),
an unpinned region persists for the lowest values of
$B/B_{\phi}$, while
the pinned region now extends above $B/B_{\phi} = 1.0$.
For higher pinning forces, the number and extent of
pinned regions grows. We illustrate this
in Fig.~\ref{fig:16} where we plot $\langle V\rangle$ versus $B/B_{\phi}$ for samples
with $F_p=0.875$ and $F_p=1.0$.
In each case, the peaks in $\langle V\rangle$
near $B/B_{\phi} = 3.0$ and $4.0$ remain present. 

\begin{figure}
\includegraphics[width=3.5in]{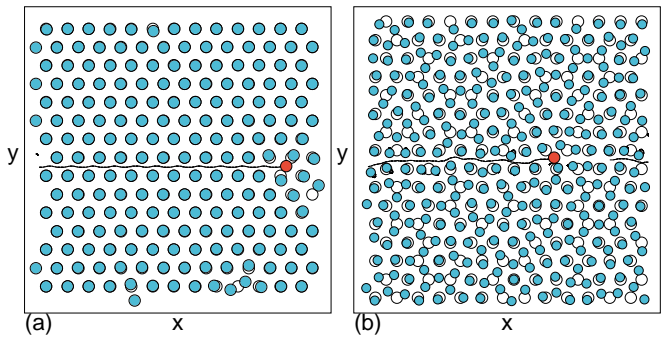}
\caption{
Image of a subsection of the superconducting vortex system with a triangular
pinning array (open circles), bulk vortices (blue circles), the driven vortex (red
circle), and the vortex trajectories for the
system in Fig.~\ref{fig:15}(a) with $F_{p} = 0.375$ and $F_D=1.0$.
The size of the pinning sites has been adjusted for clarity.
(a) Ordered flow at $B/B_{\phi} = 1.0$.
(b) Disordered flow at $B/B_{\phi} = 1.5$.
	}
\label{fig:17}
\end{figure}

In Fig.~\ref{fig:17}(a) we show the vortex positions and trajectories
for the system in Fig.~\ref{fig:15}(a) 
with $F_{p} = 0.375$, $F_D=1.0$, and $B/B_{\phi} = 1.0$.
For this filling, the background vortices form a commensurate triangular lattice and
the driven particle moves in a sinusoidal fashion through the interstitial region.
At $B/B_{\phi}=1.5$ in 
Fig.~\ref{fig:17}(b),
the system is in a disordered state and the
driven particle follows a much more random path.
We note that the velocity noise for the probe particle on a triangular pinning
array has features similar to those described above for the square pinning
array. Under commensurate conditions, there is a peak in $\langle V\rangle$, the
motion of the probe particle is periodic, and the velocity noise has a narrow band
character, while for incommensurate states, the vortex configurations are
disordered, there is no peak in $\langle V\rangle$, and the velocity noise is
broad band.

\section{Colloidal particles}

\begin{figure}
\includegraphics[width=3.5in]{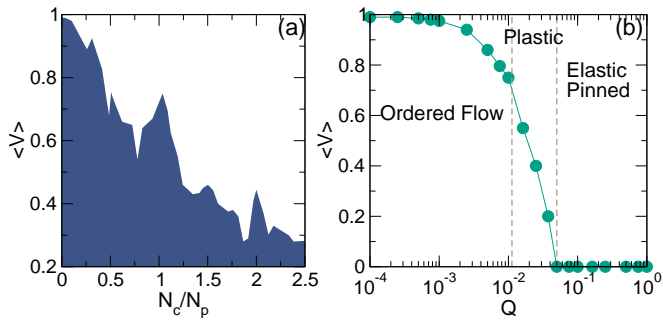}
\caption{
(a)
  $\langle V\rangle$ versus $N_{c}/N_{p}$ for a driven
  probe particle moving through a background
of colloidal particles interacting with a square pinning array, where $N_{c}$ is the number of colloids and $N_{p}$ is the number of pinning sites, for a system with colloidal charge
$Q  = 0.01$,
$F_{p} = 0.25$, and $F_{D} = 1.0$. 	
(b)	$\langle V\rangle$ versus colloidal charge $Q$ for the same system
at $N_{c}/N_{p} = 1.0$, $F_{p} = 0.25$, and $F_{D} = 1.0$.
At low $Q$, the driven particle is able to move without generating distortions in
the background lattice in the ordered flow state.
At high $Q$, the system is elastic and forms a triangular solid as
illustrated in Fig.~\ref{fig:19}(a),
and the driven particle is pinned.
For intermediate $Q$, the systems forms a partially commensurate state
of the type shown 
in Fig.~\ref{fig:19}(b). 
}
\label{fig:18}
\end{figure}

We next consider the case of a driven colloidal probe particle
moving through a background colloidal lattice
that is interacting with a square pinning array.
Here the Bessel function vortex-vortex interaction is replaced with
a screened Yukawa colloid-colloid interaction potential
of
the form $U(r) = Q\exp(-\kappa r)/r$.
In Fig.~\ref{fig:18} we plot $\langle V\rangle$ versus $N_{c}/N_{p}$, the ratio
of the number of colloids $N_c$ to the number of pinning sites $N_p$, 
for a system with a colloidal charge of $Q  = 0.01$, $F_{p} = 0.25$, and $F_{D} = 1.0$. 
Peaks in $\langle V\rangle$ appear at fillings $N_{c}/N_{p} = 1.0$ and $2.0$,
similar to what is observed in the superconducting vortex system.

\begin{figure}
\includegraphics[width=3.5in]{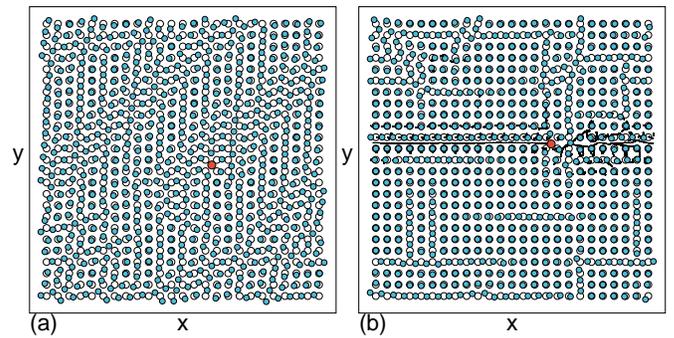}
\caption
{
Image of a subsection of the colloidal system with a square pinning array
(open circles), bulk vortices (blue circles), the driven vortex (red circle), and the
vortex trajectories for the system
in Fig~\ref{fig:18} with $F_p=0.25$, and $F_D=1.0$.
The size of the pinning sites has been adjusted for clarity.
(a) The pinned phase at $Q = 0.1$ where the
system forms a floating triangular solid.
(b) The partially commensurate phase at $Q = 0.0375$ where
the flow is plastic. 
        }
\label{fig:19}
\end{figure}

The charge on the colloidal particles can be varied, making it possible to hold the
colloid density fixed while passing from the strongly charged limit, in which
the background colloids form a stiff triangular lattice that floats above the
substrate, to the weakly charged limit, in which the pinning site locations
dominate the behavior.
In most bulk driven systems with pinning,
increasing the interactions between the particles causes the
depinning threshold to drop, and a transition can occur
from plastic depinning, where a portion of the particles are trapped by pinning sites
while the remaining particles begin to move,
to elastic depinning, where all of the particles move simultaneously and maintain
the same neighbors \cite{Reichhardt17}.
In general, the elastic depinning threshold is much lower than the plastic
depinning threshold.
For a single driven probe particle, we find the opposite behavior, in which
the driven particle becomes pinned when the background colloid interactions
become strong enough for elastic behavior to emerge.

In Fig.~\ref{fig:18}(b) we plot $\langle V\rangle$ versus $Q$
for the system in Fig.~\ref{fig:18}(a) at $N_{c}/N_{p} = 1.0$,
$F_p=0.25$, and $F_{D} = 1.0$. 
For $Q > 0.05$, the system forms a floating triangular solid
of the type illustrated in Fig.~\ref{fig:19}(a) at $Q = 0.1$. Here
the interactions between the colloidal particles are so strong that
the elastic energy of the colloidal lattice overcomes the
pinning energy of the substrate, and the background colloids float above the
substrate potential instead of sitting in the pinning sites.
Under these conditions,
the driven particle cannot tear plastically past the background colloids
and must instead push the entire system as a unit,
so $\langle V\rangle$ is
close to  zero.
For $0.01 < Q < 0.05$, the pinning energy begins to play a role and the background
colloids form
a partially commensurate  or polycrystalline solid in which
portions of the colloids are commensurate with the underlying pinning substrate
while other portions are not.
The driven particle depins plastically, generating distortions in the background
colloids as illustrated in
Fig.~\ref{fig:19}(b) for
$Q  = 0.0375$.
The plastic deformations produce
a large drag
on the driven particle,
giving a reduced $\langle V\rangle$.
At $Q = 0.01$, the background colloidal particles form a commensurate solid
similar to that shown in Fig.~\ref{fig:1}, and the
driven particle flows along a 1D channel without producing
any
plastic distortion in the lattice.
As $Q$ is decreased further,
$\langle V\rangle$ continues to increase since the drag on the driven particle
diminishes,
and the system enters
a weakly interacting limit in which the flow is along a strictly 1D channel.
At even lower $Q$, each pinning site can begin to trap multiple particles;
however,
since $F_{D}> F_{p}$, the driven particle will continue to flow even if it
occasionally passes through an empty pinning site.

\section{Discussion}
There are some limited examples
in the literature of commensurate states
that produce peaks instead of dips in the transport velocity
at
commensurate fillings.
In the work of Poccia {\it et al.} \cite{Poccia15} on superconducting vortices
moving over a square pinning array, 
at matching fields the differential resistance showed
dips at low current, indicating
enhanced pinning or reduced vortex velocities,
but at higher currents these dips became peaks,
indicating that the vortices are moving faster
at the commensurate fillings than at
the incommensurate fields.
Poccia {\it et al.} described these results in terms of
a dynamic Mott transition. 
It is possible that for low currents,
the vortex motion is dominated by the commensurate vortices trapped
at the pinning sites,
but for higher currents, there could be additional interstitial
vortices induced by the current,
and these additional
vortices could flow through
the ordered pinned vortices
in a manner similar to the motion shown in Fig.~\ref{fig:1}.
At incommensurate 
fields, there are still additional current-induced vortices when higher
currents are applied, but the vortex lattice is already disordered so the
additional drag induced by these extra vortices is not noticeable.
Jiang {\it et al.} also found a similar 
transition from dips to peaks in the
differential resistance of Nb films
with periodic pinning arrays
as a function of increasing drive \cite{Jiang04}. 
A key difference between these
experimental results and what we observe in our simulations
is that the velocity peaks
we find occur even for low drives,
indicating that the system is always behaving in an anti-commensurate fashion,
whereas in Refs.~\cite{Poccia15,Jiang04},
the velocity is enhanced at matching conditions only for high drives.

In our work we always drive along a
symmetry direction of the pinning array; however, the results could change for
driving along other directions.
In the case of the square pinning array, driving along the positive or
negative $y$ direction 
should produce the same effects as driving along the $x$ direction
due to the lattice symmetry.
We also expect similar effects to appear for driving along
other angles at which it is possible for the driven particle to follow a
1D path without encountering a pinning site, such as for driving
at $45^\circ$ from the $x$ direction.
In previous work, it was shown that particles move more easily along 
what are known as commensurate angles
on a square lattice given by
$\theta = \arctan(n/m)$ with integer $m$ and $n$\cite{Reichhardt99,Korda02},
so if the driving of the individual particle is
aligned with an incommensurate angle,
the behavior might be more similar to what is observed for
driving an individual particle over random disorder. 

In our model, the pinning is represented as
localized sites which can capture at most 
one particle.
Other types of pinning arrays also exist, such as egg carton
potentials which have no interstitial regions \cite{Reichhardt02,Brunner02}.
For such a potential, at higher fillings including $f=2$, each
potential minimum could capture two particles
which would form a dimer like state.
We expect that anti-commensuration effects would still occur in
these potentials, since at
matching conditions the
background particles would be more strongly coupled to the substrate
than to the driven particle, just as in the case of our localized pinning
sites.

Our results should be general to the broader class of systems that exhibit
commensurability effects
and that include an individually driven particle.
Our results are also relevant for understanding
the dynamics near commensurate states in the presence of some form of
doping, such as an additional particle with different properties or a different
size compared to the other particles.
Under incommensurate conditions,
the drag on these doping particles would be increased,
but the drag would be reduced in and near commensurate states.
Similar effects could occur in systems containing
multiple species that interact with each other and with
a periodic substrate. If one species
becomes more strongly coupled to the substrate,
such as at matching conditions,
the drag on the other species could be reduced.
Such behavior could be relevant for multi-component Bose-Einstein
condensates or colloidal systems
with multiple species.  

\section{Summary}

We have examined the active rheology for a driven particle moving through an
assembly of other particles in the presence of a periodic pinning
substrate. Under bulk driving, it was previously observed that
a series of commensuration effects in the form of a peak in the depinning
force or a dip in the velocity
occur when the number of particles is an integer or rational fractional
multiple of the number of pinning sites.
In this active rheology study, where only a single particle is driven,
we measure the velocity of the driven particle
at a fixed driving force and varied filling factors 
for both vortices in type-II superconductors and
colloidal particles interacting with periodic pinning arrays. 
We find an interesting anti-commensuration effect in which
the velocity of the driven particle exhibits
peaks instead of dips at the commensurate configurations,
a behavior which is the opposite of what is observed
in bulk driven systems.
Under non-matching conditions,
the system forms a disordered state
in which the driven particle entrains a greater number of background
particles and/or experiences a greater number of collisions with pinning
sites.
We show that peaks in the velocity only occur at 
matching fields for which the system forms an ordered lattice. 
Under these conditions,
the background particles form an ordered lattice which is more
strongly coupled to the substrate than to the driven particle, causing
a reduction in the drag on the driven particle.
The probe particle motion at commensurate conditions is ordered
and the velocity has a narrow band noise signature,
while at incommensurate fillings, the motion is disordered and
the noise signature becomes broad band.
In general, the probe particle has a larger velocity in the presence
of pinning than in the absence of pinning regardless of whether the
filling is commensurate or incommensurate.
The velocity-force curves indicate that
at commensuration, the depinning threshold is reduced
and the velocity in the moving phase is enhanced.
In some cases we also find dips in the probe particle velocity and
depinning force at fractional fillings such as $f=1/2$.
Within the moving phase, the velocity of the probe particle under
constant driving 
increases with increasing pinning force,
a behavior opposite from what is found in bulk driven systems. 
We show that these effects are robust for both square and triangular
pinning arrays. We also find similar effects for colloidal particles
interacting with 
a square substrate for varied filling factors. In the case of
colloidal particles, when the colloid-colloid interactions are strengthened
by increasing the charge on each colloid,
the background colloidal particles form a triangular lattice
which floats above the pinning substrate, and the probe particle becomes
pinned within this lattice,
while for lower charges and weaker colloid-colloid interactions, the probe
particle can depin and the system enters either a plastic flow or
ordered flow regime.
This is the opposite of what is observed in bulk driven systems,
where weak particle-particle interactions result in strong coupling to
the substrate and an enhanced pinning effect, while strong particle-particle
interactions produce a highly elastic particle lattice with weak
coupling to the substrate and a reduced pinning effect.
Our results should be general to a range of systems in which
particles are coupled to
periodic substrates.
The anti-commensuration effects may also explain
recent observations of vortex motion in periodic
pinning arrays,
where there is a transition from dips to peaks
in the vortex velocity at matching fields as the applied
drive is increased.
Our results 
could also be relevant to
certain systems of multiple species coupled
to a substrate in which
the drag on one species is reduced at commensurate conditions
when one species couples strongly
to the substrate while the other species does not. 

\begin{acknowledgments}
We gratefully acknowledge the support of the U.S. Department of
Energy through the LANL/LDRD program for this work.
This work was supported by the US Department of Energy through
the Los Alamos National Laboratory.  Los Alamos National Laboratory is
operated by Triad National Security, LLC, for the National Nuclear Security
Administration of the U. S. Department of Energy (Contract No. 892333218NCA000001).
\end{acknowledgments}

\bibliography{mybib}

\begin{thebibliography}{76}%
\makeatletter
\providecommand \@ifxundefined [1]{%
 \@ifx{#1\undefined}
}%
\providecommand \@ifnum [1]{%
 \ifnum #1\expandafter \@firstoftwo
 \else \expandafter \@secondoftwo
 \fi
}%
\providecommand \@ifx [1]{%
 \ifx #1\expandafter \@firstoftwo
 \else \expandafter \@secondoftwo
 \fi
}%
\providecommand \natexlab [1]{#1}%
\providecommand \enquote  [1]{``#1''}%
\providecommand \bibnamefont  [1]{#1}%
\providecommand \bibfnamefont [1]{#1}%
\providecommand \citenamefont [1]{#1}%
\providecommand \href@noop [0]{\@secondoftwo}%
\providecommand \href [0]{\begingroup \@sanitize@url \@href}%
\providecommand \@href[1]{\@@startlink{#1}\@@href}%
\providecommand \@@href[1]{\endgroup#1\@@endlink}%
\providecommand \@sanitize@url [0]{\catcode `\\12\catcode `\$12\catcode
  `\&12\catcode `\#12\catcode `\^12\catcode `\_12\catcode `\%12\relax}%
\providecommand \@@startlink[1]{}%
\providecommand \@@endlink[0]{}%
\providecommand \url  [0]{\begingroup\@sanitize@url \@url }%
\providecommand \@url [1]{\endgroup\@href {#1}{\urlprefix }}%
\providecommand \urlprefix  [0]{URL }%
\providecommand \Eprint [0]{\href }%
\providecommand \doibase [0]{http://dx.doi.org/}%
\providecommand \selectlanguage [0]{\@gobble}%
\providecommand \bibinfo  [0]{\@secondoftwo}%
\providecommand \bibfield  [0]{\@secondoftwo}%
\providecommand \translation [1]{[#1]}%
\providecommand \BibitemOpen [0]{}%
\providecommand \bibitemStop [0]{}%
\providecommand \bibitemNoStop [0]{.\EOS\space}%
\providecommand \EOS [0]{\spacefactor3000\relax}%
\providecommand \BibitemShut  [1]{\csname bibitem#1\endcsname}%
\let\auto@bib@innerbib\@empty
%</preamble>
\bibitem [{\citenamefont {Bak}(1982)}]{Bak82}%
  \BibitemOpen
  \bibfield  {author} {\bibinfo {author} {\bibfnamefont {P.}~\bibnamefont
  {Bak}},\ }\bibfield  {title} {\enquote {\bibinfo {title} {Commensurate
  phases, incommensurate phases and the devil's staricase},}\ }\href {\doibase
  10.1088/0034-4885/45/6/001} {\bibfield  {journal} {\bibinfo  {journal} {Rep.
  Prog. Phys.}\ }\textbf {\bibinfo {volume} {45}},\ \bibinfo {pages} {587--629}
  (\bibinfo {year} {1982})}\BibitemShut {NoStop}%
\bibitem [{\citenamefont {Harada}\ \emph {et~al.}(1996)\citenamefont {Harada},
  \citenamefont {Kamimura}, \citenamefont {Kasai}, \citenamefont {Matsuda},
  \citenamefont {Tonomura},\ and\ \citenamefont {Moshchalkov}}]{Harada96}%
  \BibitemOpen
  \bibfield  {author} {\bibinfo {author} {\bibfnamefont {K.}~\bibnamefont
  {Harada}}, \bibinfo {author} {\bibfnamefont {O.}~\bibnamefont {Kamimura}},
  \bibinfo {author} {\bibfnamefont {H.}~\bibnamefont {Kasai}}, \bibinfo
  {author} {\bibfnamefont {T.}~\bibnamefont {Matsuda}}, \bibinfo {author}
  {\bibfnamefont {A.}~\bibnamefont {Tonomura}}, \ and\ \bibinfo {author}
  {\bibfnamefont {V.~V.}\ \bibnamefont {Moshchalkov}},\ }\bibfield  {title}
  {\enquote {\bibinfo {title} {Direct observation of vortex dynamics in
  superconducting films with regular arrays of defects},}\ }\href {\doibase
  10.1126/science.274.5290.1167} {\bibfield  {journal} {\bibinfo  {journal}
  {Science}\ }\textbf {\bibinfo {volume} {274}},\ \bibinfo {pages} {1167--1170}
  (\bibinfo {year} {1996})}\BibitemShut {NoStop}%
\bibitem [{\citenamefont {Hu}\ and\ \citenamefont {Westervelt}(1997)}]{Hu97}%
  \BibitemOpen
  \bibfield  {author} {\bibinfo {author} {\bibfnamefont {J.}~\bibnamefont
  {Hu}}\ and\ \bibinfo {author} {\bibfnamefont {R.~M.}\ \bibnamefont
  {Westervelt}},\ }\bibfield  {title} {\enquote {\bibinfo {title}
  {Commensurate-incommensurate transitions in magnetic bubble arrays with
  periodic line pinning},}\ }\href {\doibase 10.1103/PhysRevB.55.771}
  {\bibfield  {journal} {\bibinfo  {journal} {Phys. Rev. B}\ }\textbf {\bibinfo
  {volume} {55}},\ \bibinfo {pages} {771--774} (\bibinfo {year}
  {1997})}\BibitemShut {NoStop}%
\bibitem [{\citenamefont {Mangold}\ \emph {et~al.}(2003)\citenamefont
  {Mangold}, \citenamefont {Leiderer},\ and\ \citenamefont
  {Bechinger}}]{Mangold03}%
  \BibitemOpen
  \bibfield  {author} {\bibinfo {author} {\bibfnamefont {K.}~\bibnamefont
  {Mangold}}, \bibinfo {author} {\bibfnamefont {P.}~\bibnamefont {Leiderer}}, \
  and\ \bibinfo {author} {\bibfnamefont {C.}~\bibnamefont {Bechinger}},\
  }\bibfield  {title} {\enquote {\bibinfo {title} {Phase transitions of
  colloidal monolayers in periodic pinning arrays},}\ }\href {\doibase
  10.1103/PhysRevLett.90.158302} {\bibfield  {journal} {\bibinfo  {journal}
  {Phys. Rev. Lett.}\ }\textbf {\bibinfo {volume} {90}},\ \bibinfo {pages}
  {158302} (\bibinfo {year} {2003})}\BibitemShut {NoStop}%
\bibitem [{\citenamefont {Tung}\ \emph {et~al.}(2006)\citenamefont {Tung},
  \citenamefont {Schweikhard},\ and\ \citenamefont {Cornell}}]{Tung06}%
  \BibitemOpen
  \bibfield  {author} {\bibinfo {author} {\bibfnamefont {S.}~\bibnamefont
  {Tung}}, \bibinfo {author} {\bibfnamefont {V.}~\bibnamefont {Schweikhard}}, \
  and\ \bibinfo {author} {\bibfnamefont {E.~A.}\ \bibnamefont {Cornell}},\
  }\bibfield  {title} {\enquote {\bibinfo {title} {Observation of vortex
  pinning in {Bose-E}instein condensates},}\ }\href {\doibase
  10.1103/PhysRevLett.97.240402} {\bibfield  {journal} {\bibinfo  {journal}
  {Phys. Rev. Lett.}\ }\textbf {\bibinfo {volume} {97}},\ \bibinfo {pages}
  {240402} (\bibinfo {year} {2006})}\BibitemShut {NoStop}%
\bibitem [{\citenamefont {McDermott}\ \emph
  {et~al.}(2013{\natexlab{a}})\citenamefont {McDermott}, \citenamefont
  {Amelang}, \citenamefont {Lopatina}, \citenamefont {Reichhardt},\ and\
  \citenamefont {Reichhardt}}]{McDermott13}%
  \BibitemOpen
  \bibfield  {author} {\bibinfo {author} {\bibfnamefont {D.}~\bibnamefont
  {McDermott}}, \bibinfo {author} {\bibfnamefont {J.}~\bibnamefont {Amelang}},
  \bibinfo {author} {\bibfnamefont {L.~M.}\ \bibnamefont {Lopatina}}, \bibinfo
  {author} {\bibfnamefont {C.~J.~Olson}\ \bibnamefont {Reichhardt}}, \ and\
  \bibinfo {author} {\bibfnamefont {C.}~\bibnamefont {Reichhardt}},\ }\bibfield
   {title} {\enquote {\bibinfo {title} {Domain and stripe formation between
  hexagonal and square ordered fillings of colloidal particles on periodic
  pinning substrates},}\ }\href {\doibase 10.1039/c3sm27652j} {\bibfield
  {journal} {\bibinfo  {journal} {Soft Matter}\ }\textbf {\bibinfo {volume}
  {9}},\ \bibinfo {pages} {4607--4613} (\bibinfo {year}
  {2013}{\natexlab{a}})}\BibitemShut {NoStop}%
\bibitem [{\citenamefont {Vanossi}\ \emph {et~al.}(2013)\citenamefont
  {Vanossi}, \citenamefont {Manini}, \citenamefont {Urbakh}, \citenamefont
  {Zapperi},\ and\ \citenamefont {Tosatti}}]{Vanossi13}%
  \BibitemOpen
  \bibfield  {author} {\bibinfo {author} {\bibfnamefont {A.}~\bibnamefont
  {Vanossi}}, \bibinfo {author} {\bibfnamefont {N.}~\bibnamefont {Manini}},
  \bibinfo {author} {\bibfnamefont {M.}~\bibnamefont {Urbakh}}, \bibinfo
  {author} {\bibfnamefont {S.}~\bibnamefont {Zapperi}}, \ and\ \bibinfo
  {author} {\bibfnamefont {E.}~\bibnamefont {Tosatti}},\ }\bibfield  {title}
  {\enquote {\bibinfo {title} {Colloquium: Modeling friction: From nanoscale to
  mesoscale},}\ }\href {\doibase 10.1103/RevModPhys.85.529} {\bibfield
  {journal} {\bibinfo  {journal} {Rev. Mod. Phys.}\ }\textbf {\bibinfo {volume}
  {85}},\ \bibinfo {pages} {529--552} (\bibinfo {year} {2013})}\BibitemShut
  {NoStop}%
\bibitem [{\citenamefont {Bohlein}\ \emph {et~al.}(2012)\citenamefont
  {Bohlein}, \citenamefont {Mikhael},\ and\ \citenamefont
  {Bechinger}}]{Bohlein12}%
  \BibitemOpen
  \bibfield  {author} {\bibinfo {author} {\bibfnamefont {T.}~\bibnamefont
  {Bohlein}}, \bibinfo {author} {\bibfnamefont {J.}~\bibnamefont {Mikhael}}, \
  and\ \bibinfo {author} {\bibfnamefont {C.}~\bibnamefont {Bechinger}},\
  }\bibfield  {title} {\enquote {\bibinfo {title} {Observation of kinks and
  antikinks in colloidal monolayers driven across ordered surfaces},}\ }\href
  {\doibase 10.1038/NMAT3204} {\bibfield  {journal} {\bibinfo  {journal}
  {Nature Mater.}\ }\textbf {\bibinfo {volume} {11}},\ \bibinfo {pages}
  {126--130} (\bibinfo {year} {2012})}\BibitemShut {NoStop}%
\bibitem [{\citenamefont {Reichhardt}\ and\ \citenamefont
  {Reichhardt}(2017)}]{Reichhardt17}%
  \BibitemOpen
  \bibfield  {author} {\bibinfo {author} {\bibfnamefont {C.}~\bibnamefont
  {Reichhardt}}\ and\ \bibinfo {author} {\bibfnamefont {C.~J.~Olson}\
  \bibnamefont {Reichhardt}},\ }\bibfield  {title} {\enquote {\bibinfo {title}
  {Depinning and nonequilibrium dynamic phases of particle assemblies driven
  over random and ordered substrates: a review},}\ }\href {\doibase
  10.1088/1361-6633/80/2/026501} {\bibfield  {journal} {\bibinfo  {journal}
  {Rep. Prog. Phys.}\ }\textbf {\bibinfo {volume} {80}},\ \bibinfo {pages}
  {026501} (\bibinfo {year} {2017})}\BibitemShut {NoStop}%
\bibitem [{\citenamefont {Gutierrez}\ \emph {et~al.}(2009)\citenamefont
  {Gutierrez}, \citenamefont {Silhanek}, \citenamefont {Van~de Vondel},
  \citenamefont {Gillijns},\ and\ \citenamefont {Moshchalkov}}]{Gutierrez09}%
  \BibitemOpen
  \bibfield  {author} {\bibinfo {author} {\bibfnamefont {J.}~\bibnamefont
  {Gutierrez}}, \bibinfo {author} {\bibfnamefont {A.~V.}\ \bibnamefont
  {Silhanek}}, \bibinfo {author} {\bibfnamefont {J.}~\bibnamefont {Van~de
  Vondel}}, \bibinfo {author} {\bibfnamefont {W.}~\bibnamefont {Gillijns}}, \
  and\ \bibinfo {author} {\bibfnamefont {V.~V.}\ \bibnamefont {Moshchalkov}},\
  }\bibfield  {title} {\enquote {\bibinfo {title} {Transition from turbulent to
  nearly laminar vortex flow in superconductors with periodic pinning},}\
  }\href {\doibase 10.1103/PhysRevB.80.140514} {\bibfield  {journal} {\bibinfo
  {journal} {Phys. Rev. B}\ }\textbf {\bibinfo {volume} {80}},\ \bibinfo
  {pages} {140514} (\bibinfo {year} {2009})}\BibitemShut {NoStop}%
\bibitem [{\citenamefont {Avci}\ \emph {et~al.}(2010)\citenamefont {Avci},
  \citenamefont {Xiao}, \citenamefont {Hua}, \citenamefont {Imre},
  \citenamefont {Divan}, \citenamefont {Pearson}, \citenamefont {Welp},
  \citenamefont {Kwok},\ and\ \citenamefont {Crabtree}}]{Avci10}%
  \BibitemOpen
  \bibfield  {author} {\bibinfo {author} {\bibfnamefont {S.}~\bibnamefont
  {Avci}}, \bibinfo {author} {\bibfnamefont {Z.~L.}\ \bibnamefont {Xiao}},
  \bibinfo {author} {\bibfnamefont {J.}~\bibnamefont {Hua}}, \bibinfo {author}
  {\bibfnamefont {A.}~\bibnamefont {Imre}}, \bibinfo {author} {\bibfnamefont
  {R.}~\bibnamefont {Divan}}, \bibinfo {author} {\bibfnamefont
  {J.}~\bibnamefont {Pearson}}, \bibinfo {author} {\bibfnamefont
  {U.}~\bibnamefont {Welp}}, \bibinfo {author} {\bibfnamefont {W.~K.}\
  \bibnamefont {Kwok}}, \ and\ \bibinfo {author} {\bibfnamefont {G.~W.}\
  \bibnamefont {Crabtree}},\ }\bibfield  {title} {\enquote {\bibinfo {title}
  {Matching effect and dynamic phases of vortex matter in
  {Bi$_2$Sr$_2$CaCu$_2$O$_8$} nanoribbon with a periodic array of holes},}\
  }\href {\doibase 10.1063/1.3473783} {\bibfield  {journal} {\bibinfo
  {journal} {Appl. Phys. Lett.}\ }\textbf {\bibinfo {volume} {97}},\ \bibinfo
  {pages} {042511} (\bibinfo {year} {2010})}\BibitemShut {NoStop}%
\bibitem [{\citenamefont {Vanossi}\ \emph {et~al.}(2012)\citenamefont
  {Vanossi}, \citenamefont {Manini},\ and\ \citenamefont
  {Tosatti}}]{Vanossi12}%
  \BibitemOpen
  \bibfield  {author} {\bibinfo {author} {\bibfnamefont {A.}~\bibnamefont
  {Vanossi}}, \bibinfo {author} {\bibfnamefont {N.}~\bibnamefont {Manini}}, \
  and\ \bibinfo {author} {\bibfnamefont {E.}~\bibnamefont {Tosatti}},\
  }\bibfield  {title} {\enquote {\bibinfo {title} {Static and dynamic friction
  in sliding colloidal monolayers},}\ }\href {\doibase 10.1073/pnas.1213930109}
  {\bibfield  {journal} {\bibinfo  {journal} {Proc. Natl. Acad. Sci. (USA)}\
  }\textbf {\bibinfo {volume} {109}},\ \bibinfo {pages} {16429--16433}
  (\bibinfo {year} {2012})}\BibitemShut {NoStop}%
\bibitem [{\citenamefont {Daldini}\ \emph {et~al.}(1974)\citenamefont
  {Daldini}, \citenamefont {Martinoli}, \citenamefont {Olsen},\ and\
  \citenamefont {Berner}}]{Daldini74}%
  \BibitemOpen
  \bibfield  {author} {\bibinfo {author} {\bibfnamefont {O.}~\bibnamefont
  {Daldini}}, \bibinfo {author} {\bibfnamefont {P.}~\bibnamefont {Martinoli}},
  \bibinfo {author} {\bibfnamefont {J.~L.}\ \bibnamefont {Olsen}}, \ and\
  \bibinfo {author} {\bibfnamefont {G.}~\bibnamefont {Berner}},\ }\bibfield
  {title} {\enquote {\bibinfo {title} {Vortex-line pinning by thickness
  modulation of superconducting films},}\ }\href {\doibase
  10.1103/PhysRevLett.32.218} {\bibfield  {journal} {\bibinfo  {journal} {Phys.
  Rev. Lett.}\ }\textbf {\bibinfo {volume} {32}},\ \bibinfo {pages} {218--221}
  (\bibinfo {year} {1974})}\BibitemShut {NoStop}%
\bibitem [{\citenamefont {Dobrovolskiy}\ \emph {et~al.}(2012)\citenamefont
  {Dobrovolskiy}, \citenamefont {Begun}, \citenamefont {Huth},\ and\
  \citenamefont {Shklovskij}}]{Dobrovolskiy12}%
  \BibitemOpen
  \bibfield  {author} {\bibinfo {author} {\bibfnamefont {O.~V.}\ \bibnamefont
  {Dobrovolskiy}}, \bibinfo {author} {\bibfnamefont {E.}~\bibnamefont {Begun}},
  \bibinfo {author} {\bibfnamefont {M.}~\bibnamefont {Huth}}, \ and\ \bibinfo
  {author} {\bibfnamefont {V.~A.}\ \bibnamefont {Shklovskij}},\ }\bibfield
  {title} {\enquote {\bibinfo {title} {Electrical transport and pinning
  properties of {Nb} thin films patterned with focused ion beam-milled
  washboard nanostructures},}\ }\href {\doibase 10.1088/1367-2630/14/11/113027}
  {\bibfield  {journal} {\bibinfo  {journal} {New J. Phys.}\ }\textbf {\bibinfo
  {volume} {14}},\ \bibinfo {pages} {113027} (\bibinfo {year}
  {2012})}\BibitemShut {NoStop}%
\bibitem [{\citenamefont {Le~Thien}\ \emph {et~al.}(2016)\citenamefont
  {Le~Thien}, \citenamefont {McDermott}, \citenamefont {Olson~Reichhardt},\
  and\ \citenamefont {Reichhardt}}]{LeThien16}%
  \BibitemOpen
  \bibfield  {author} {\bibinfo {author} {\bibfnamefont {Q.}~\bibnamefont
  {Le~Thien}}, \bibinfo {author} {\bibfnamefont {D.}~\bibnamefont {McDermott}},
  \bibinfo {author} {\bibfnamefont {C.~J.}\ \bibnamefont {Olson~Reichhardt}}, \
  and\ \bibinfo {author} {\bibfnamefont {C.}~\bibnamefont {Reichhardt}},\
  }\bibfield  {title} {\enquote {\bibinfo {title} {Orientational ordering,
  buckling, and dynamic transitions for vortices interacting with a periodic
  quasi-one-dimensional substrate},}\ }\href {\doibase
  10.1103/PhysRevB.93.014504} {\bibfield  {journal} {\bibinfo  {journal} {Phys.
  Rev. B}\ }\textbf {\bibinfo {volume} {93}},\ \bibinfo {pages} {014504}
  (\bibinfo {year} {2016})}\BibitemShut {NoStop}%
\bibitem [{\citenamefont {Baert}\ \emph {et~al.}(1995)\citenamefont {Baert},
  \citenamefont {Metlushko}, \citenamefont {Jonckheere}, \citenamefont
  {Moshchalkov},\ and\ \citenamefont {Bruynseraede}}]{Baert95}%
  \BibitemOpen
  \bibfield  {author} {\bibinfo {author} {\bibfnamefont {M.}~\bibnamefont
  {Baert}}, \bibinfo {author} {\bibfnamefont {V.~V.}\ \bibnamefont
  {Metlushko}}, \bibinfo {author} {\bibfnamefont {R.}~\bibnamefont
  {Jonckheere}}, \bibinfo {author} {\bibfnamefont {V.~V.}\ \bibnamefont
  {Moshchalkov}}, \ and\ \bibinfo {author} {\bibfnamefont {Y.}~\bibnamefont
  {Bruynseraede}},\ }\bibfield  {title} {\enquote {\bibinfo {title} {Composite
  flux-line lattices stabilized in superconducting films by a regular array of
  artificial defects},}\ }\href {\doibase 10.1103/PhysRevLett.74.3269}
  {\bibfield  {journal} {\bibinfo  {journal} {Phys. Rev. Lett.}\ }\textbf
  {\bibinfo {volume} {74}},\ \bibinfo {pages} {3269--3272} (\bibinfo {year}
  {1995})}\BibitemShut {NoStop}%
\bibitem [{\citenamefont {Reichhardt}\ \emph
  {et~al.}(1998{\natexlab{a}})\citenamefont {Reichhardt}, \citenamefont
  {Olson},\ and\ \citenamefont {Nori}}]{Reichhardt98a}%
  \BibitemOpen
  \bibfield  {author} {\bibinfo {author} {\bibfnamefont {C.}~\bibnamefont
  {Reichhardt}}, \bibinfo {author} {\bibfnamefont {C.~J.}\ \bibnamefont
  {Olson}}, \ and\ \bibinfo {author} {\bibfnamefont {F.}~\bibnamefont {Nori}},\
  }\bibfield  {title} {\enquote {\bibinfo {title} {Commensurate and
  incommensurate vortex states in superconductors with periodic pinning
  arrays},}\ }\href {\doibase 10.1103/PhysRevB.57.7937} {\bibfield  {journal}
  {\bibinfo  {journal} {Phys. Rev. B}\ }\textbf {\bibinfo {volume} {57}},\
  \bibinfo {pages} {7937--7943} (\bibinfo {year}
  {1998}{\natexlab{a}})}\BibitemShut {NoStop}%
\bibitem [{\citenamefont {Reichhardt}\ and\ \citenamefont
  {Olson}(2002{\natexlab{a}})}]{Reichhardt02}%
  \BibitemOpen
  \bibfield  {author} {\bibinfo {author} {\bibfnamefont {C.}~\bibnamefont
  {Reichhardt}}\ and\ \bibinfo {author} {\bibfnamefont {C.~J.}\ \bibnamefont
  {Olson}},\ }\bibfield  {title} {\enquote {\bibinfo {title} {Colloidal
  dynamics on disordered substrates},}\ }\href {\doibase
  10.1103/PhysRevLett.89.078301} {\bibfield  {journal} {\bibinfo  {journal}
  {Phys. Rev. Lett.}\ }\textbf {\bibinfo {volume} {89}},\ \bibinfo {pages}
  {078301} (\bibinfo {year} {2002}{\natexlab{a}})}\BibitemShut {NoStop}%
\bibitem [{\citenamefont {Brunner}\ and\ \citenamefont
  {Bechinger}(2002)}]{Brunner02}%
  \BibitemOpen
  \bibfield  {author} {\bibinfo {author} {\bibfnamefont {M.}~\bibnamefont
  {Brunner}}\ and\ \bibinfo {author} {\bibfnamefont {C.}~\bibnamefont
  {Bechinger}},\ }\bibfield  {title} {\enquote {\bibinfo {title} {Phase
  behavior of colloidal molecular crystals on triangular light lattices},}\
  }\href {\doibase 10.1103/PhysRevLett.88.248302} {\bibfield  {journal}
  {\bibinfo  {journal} {Phys. Rev. Lett.}\ }\textbf {\bibinfo {volume} {88}},\
  \bibinfo {pages} {248302} (\bibinfo {year} {2002})}\BibitemShut {NoStop}%
\bibitem [{\citenamefont {Mart\'{\i}n}\ \emph {et~al.}(1997)\citenamefont
  {Mart\'{\i}n}, \citenamefont {V\'elez}, \citenamefont {Nogu\'es},\ and\
  \citenamefont {Schuller}}]{Martin97}%
  \BibitemOpen
  \bibfield  {author} {\bibinfo {author} {\bibfnamefont {J.~I.}\ \bibnamefont
  {Mart\'{\i}n}}, \bibinfo {author} {\bibfnamefont {M.}~\bibnamefont
  {V\'elez}}, \bibinfo {author} {\bibfnamefont {J.}~\bibnamefont {Nogu\'es}}, \
  and\ \bibinfo {author} {\bibfnamefont {I.~K.}\ \bibnamefont {Schuller}},\
  }\bibfield  {title} {\enquote {\bibinfo {title} {Flux pinning in a
  superconductor by an array of submicrometer magnetic dots},}\ }\href
  {\doibase 10.1103/PhysRevLett.79.1929} {\bibfield  {journal} {\bibinfo
  {journal} {Phys. Rev. Lett.}\ }\textbf {\bibinfo {volume} {79}},\ \bibinfo
  {pages} {1929--1932} (\bibinfo {year} {1997})}\BibitemShut {NoStop}%
\bibitem [{\citenamefont {Grigorenko}\ \emph {et~al.}(2003)\citenamefont
  {Grigorenko}, \citenamefont {Bending}, \citenamefont {Van~Bael},
  \citenamefont {Lange}, \citenamefont {Moshchalkov}, \citenamefont {Fangohr},\
  and\ \citenamefont {de~Groot}}]{Grigorenko03}%
  \BibitemOpen
  \bibfield  {author} {\bibinfo {author} {\bibfnamefont {A.~N.}\ \bibnamefont
  {Grigorenko}}, \bibinfo {author} {\bibfnamefont {S.~J.}\ \bibnamefont
  {Bending}}, \bibinfo {author} {\bibfnamefont {M.~J.}\ \bibnamefont
  {Van~Bael}}, \bibinfo {author} {\bibfnamefont {M.}~\bibnamefont {Lange}},
  \bibinfo {author} {\bibfnamefont {V.~V.}\ \bibnamefont {Moshchalkov}},
  \bibinfo {author} {\bibfnamefont {H.}~\bibnamefont {Fangohr}}, \ and\
  \bibinfo {author} {\bibfnamefont {P.~A.~J.}\ \bibnamefont {de~Groot}},\
  }\bibfield  {title} {\enquote {\bibinfo {title} {Symmetry locking and
  commensurate vortex domain formation in periodic pinning arrays},}\ }\href
  {\doibase 10.1103/PhysRevLett.90.237001} {\bibfield  {journal} {\bibinfo
  {journal} {Phys. Rev. Lett.}\ }\textbf {\bibinfo {volume} {90}},\ \bibinfo
  {pages} {237001} (\bibinfo {year} {2003})}\BibitemShut {NoStop}%
\bibitem [{\citenamefont {Berdiyorov}\ \emph {et~al.}(2006)\citenamefont
  {Berdiyorov}, \citenamefont {Milo\v{s}evi\'{c}},\ and\ \citenamefont
  {Peeters}}]{Berdiyorov06}%
  \BibitemOpen
  \bibfield  {author} {\bibinfo {author} {\bibfnamefont {G.~R.}\ \bibnamefont
  {Berdiyorov}}, \bibinfo {author} {\bibfnamefont {M.~V.}\ \bibnamefont
  {Milo\v{s}evi\'{c}}}, \ and\ \bibinfo {author} {\bibfnamefont {F.~M.}\
  \bibnamefont {Peeters}},\ }\bibfield  {title} {\enquote {\bibinfo {title}
  {Novel commensurability effects in superconducting films with antidot
  arrays},}\ }\href {\doibase 10.1103/PhysRevLett.96.207001} {\bibfield
  {journal} {\bibinfo  {journal} {Phys. Rev. Lett.}\ }\textbf {\bibinfo
  {volume} {96}},\ \bibinfo {pages} {207001} (\bibinfo {year}
  {2006})}\BibitemShut {NoStop}%
\bibitem [{\citenamefont {Latimer}\ \emph {et~al.}(2013)\citenamefont
  {Latimer}, \citenamefont {Berdiyorov}, \citenamefont {Xiao}, \citenamefont
  {Peeters},\ and\ \citenamefont {Kwok}}]{Latimer13}%
  \BibitemOpen
  \bibfield  {author} {\bibinfo {author} {\bibfnamefont {M.~L.}\ \bibnamefont
  {Latimer}}, \bibinfo {author} {\bibfnamefont {G.~R.}\ \bibnamefont
  {Berdiyorov}}, \bibinfo {author} {\bibfnamefont {Z.~L.}\ \bibnamefont
  {Xiao}}, \bibinfo {author} {\bibfnamefont {F.~M.}\ \bibnamefont {Peeters}}, \
  and\ \bibinfo {author} {\bibfnamefont {W.~K.}\ \bibnamefont {Kwok}},\
  }\bibfield  {title} {\enquote {\bibinfo {title} {Realization of artificial
  ice systems for magnetic vortices in a superconducting {MoGe} thin film with
  patterned nanostructures},}\ }\href {\doibase 10.1103/PhysRevLett.111.067001}
  {\bibfield  {journal} {\bibinfo  {journal} {Phys. Rev. Lett.}\ }\textbf
  {\bibinfo {volume} {111}},\ \bibinfo {pages} {067001} (\bibinfo {year}
  {2013})}\BibitemShut {NoStop}%
\bibitem [{\citenamefont {Reichhardt}\ and\ \citenamefont
  {Gr\o{}nbech-Jensen}(2001)}]{Reichhardt01a}%
  \BibitemOpen
  \bibfield  {author} {\bibinfo {author} {\bibfnamefont {C.}~\bibnamefont
  {Reichhardt}}\ and\ \bibinfo {author} {\bibfnamefont {N.}~\bibnamefont
  {Gr\o{}nbech-Jensen}},\ }\bibfield  {title} {\enquote {\bibinfo {title}
  {Critical currents and vortex states at fractional matching fields in
  superconductors with periodic pinning},}\ }\href {\doibase
  10.1103/PhysRevB.63.054510} {\bibfield  {journal} {\bibinfo  {journal} {Phys.
  Rev. B}\ }\textbf {\bibinfo {volume} {63}},\ \bibinfo {pages} {054510}
  (\bibinfo {year} {2001})}\BibitemShut {NoStop}%
\bibitem [{\citenamefont {Trastoy}\ \emph {et~al.}(2014)\citenamefont
  {Trastoy}, \citenamefont {Malnou}, \citenamefont {Ulysse}, \citenamefont
  {Bernard}, \citenamefont {Bergeal}, \citenamefont {Faini}, \citenamefont
  {Lesueur}, \citenamefont {Briatico},\ and\ \citenamefont
  {Villegas}}]{Trastoy14}%
  \BibitemOpen
  \bibfield  {author} {\bibinfo {author} {\bibfnamefont {J.}~\bibnamefont
  {Trastoy}}, \bibinfo {author} {\bibfnamefont {M.}~\bibnamefont {Malnou}},
  \bibinfo {author} {\bibfnamefont {C.}~\bibnamefont {Ulysse}}, \bibinfo
  {author} {\bibfnamefont {R.}~\bibnamefont {Bernard}}, \bibinfo {author}
  {\bibfnamefont {N.}~\bibnamefont {Bergeal}}, \bibinfo {author} {\bibfnamefont
  {G.}~\bibnamefont {Faini}}, \bibinfo {author} {\bibfnamefont
  {J.}~\bibnamefont {Lesueur}}, \bibinfo {author} {\bibfnamefont
  {J.}~\bibnamefont {Briatico}}, \ and\ \bibinfo {author} {\bibfnamefont
  {J.~E.}\ \bibnamefont {Villegas}},\ }\bibfield  {title} {\enquote {\bibinfo
  {title} {Freezing and thawing of artificial ice by thermal switching of
  geometric frustration in magnetic flux lattices},}\ }\href {\doibase
  10.1038/NNANO.2014.158} {\bibfield  {journal} {\bibinfo  {journal} {Nature
  Nanotechnol.}\ }\textbf {\bibinfo {volume} {9}},\ \bibinfo {pages} {710--715}
  (\bibinfo {year} {2014})}\BibitemShut {NoStop}%
\bibitem [{\citenamefont {Sadovskyy}\ \emph {et~al.}(2017)\citenamefont
  {Sadovskyy}, \citenamefont {Wang}, \citenamefont {Xiao}, \citenamefont
  {Kwok},\ and\ \citenamefont {Glatz}}]{Sadovskyy17}%
  \BibitemOpen
  \bibfield  {author} {\bibinfo {author} {\bibfnamefont {I.~A.}\ \bibnamefont
  {Sadovskyy}}, \bibinfo {author} {\bibfnamefont {Y.~L.}\ \bibnamefont {Wang}},
  \bibinfo {author} {\bibfnamefont {Z.-L.}\ \bibnamefont {Xiao}}, \bibinfo
  {author} {\bibfnamefont {W.-K.}\ \bibnamefont {Kwok}}, \ and\ \bibinfo
  {author} {\bibfnamefont {A.}~\bibnamefont {Glatz}},\ }\bibfield  {title}
  {\enquote {\bibinfo {title} {Effect of hexagonal patterned arrays and defect
  geometry on the critical current of superconducting films},}\ }\href
  {\doibase 10.1103/PhysRevB.95.075303} {\bibfield  {journal} {\bibinfo
  {journal} {Phys. Rev. B}\ }\textbf {\bibinfo {volume} {95}},\ \bibinfo
  {pages} {075303} (\bibinfo {year} {2017})}\BibitemShut {NoStop}%
\bibitem [{\citenamefont {Wang}\ \emph {et~al.}(2018)\citenamefont {Wang},
  \citenamefont {Ma}, \citenamefont {Xu}, \citenamefont {Xiao}, \citenamefont
  {Snezhko}, \citenamefont {Divan}, \citenamefont {Ocola}, \citenamefont
  {Pearson}, \citenamefont {J{\' a}nko},\ and\ \citenamefont {Kwok}}]{Wang18}%
  \BibitemOpen
  \bibfield  {author} {\bibinfo {author} {\bibfnamefont {Y.-L.}\ \bibnamefont
  {Wang}}, \bibinfo {author} {\bibfnamefont {X.}~\bibnamefont {Ma}}, \bibinfo
  {author} {\bibfnamefont {J.}~\bibnamefont {Xu}}, \bibinfo {author}
  {\bibfnamefont {Z.-L.}\ \bibnamefont {Xiao}}, \bibinfo {author}
  {\bibfnamefont {A.}~\bibnamefont {Snezhko}}, \bibinfo {author} {\bibfnamefont
  {R.}~\bibnamefont {Divan}}, \bibinfo {author} {\bibfnamefont {L.~E.}\
  \bibnamefont {Ocola}}, \bibinfo {author} {\bibfnamefont {J.~E.}\ \bibnamefont
  {Pearson}}, \bibinfo {author} {\bibfnamefont {B.}~\bibnamefont {J{\' a}nko}},
  \ and\ \bibinfo {author} {\bibfnamefont {W.-K.}\ \bibnamefont {Kwok}},\
  }\bibfield  {title} {\enquote {\bibinfo {title} {Switchable geometric
  frustration in an artificial-spin-ice-superconductor heterosystem},}\ }\href
  {\doibase 10.1038/s41565-018-0162-7} {\bibfield  {journal} {\bibinfo
  {journal} {Nature Nanotechnol.}\ }\textbf {\bibinfo {volume} {13}},\ \bibinfo
  {pages} {560} (\bibinfo {year} {2018})}\BibitemShut {NoStop}%
\bibitem [{\citenamefont {Reichhardt}\ and\ \citenamefont
  {Olson}(2002{\natexlab{b}})}]{Reichhardt02a}%
  \BibitemOpen
  \bibfield  {author} {\bibinfo {author} {\bibfnamefont {C.}~\bibnamefont
  {Reichhardt}}\ and\ \bibinfo {author} {\bibfnamefont {C.~J.}\ \bibnamefont
  {Olson}},\ }\bibfield  {title} {\enquote {\bibinfo {title} {Novel colloidal
  crystalline states on two-dimensional periodic substrates},}\ }\href
  {\doibase 10.1103/PhysRevLett.88.248301} {\bibfield  {journal} {\bibinfo
  {journal} {Phys. Rev. Lett.}\ }\textbf {\bibinfo {volume} {88}},\ \bibinfo
  {pages} {248301} (\bibinfo {year} {2002}{\natexlab{b}})}\BibitemShut
  {NoStop}%
\bibitem [{\citenamefont {Brazda}\ \emph {et~al.}(2018)\citenamefont {Brazda},
  \citenamefont {Silva}, \citenamefont {Manini}, \citenamefont {Vanossi},
  \citenamefont {Guerra}, \citenamefont {Tosatti},\ and\ \citenamefont
  {Bechinger}}]{Brazda18}%
  \BibitemOpen
  \bibfield  {author} {\bibinfo {author} {\bibfnamefont {T.}~\bibnamefont
  {Brazda}}, \bibinfo {author} {\bibfnamefont {A.}~\bibnamefont {Silva}},
  \bibinfo {author} {\bibfnamefont {N.}~\bibnamefont {Manini}}, \bibinfo
  {author} {\bibfnamefont {A.}~\bibnamefont {Vanossi}}, \bibinfo {author}
  {\bibfnamefont {R.}~\bibnamefont {Guerra}}, \bibinfo {author} {\bibfnamefont
  {E.}~\bibnamefont {Tosatti}}, \ and\ \bibinfo {author} {\bibfnamefont
  {C.}~\bibnamefont {Bechinger}},\ }\bibfield  {title} {\enquote {\bibinfo
  {title} {Experimental observation of the {A}ubry transition in
  two-dimensional colloidal monolayers},}\ }\href {\doibase
  10.1103/PhysRevX.8.011050} {\bibfield  {journal} {\bibinfo  {journal} {Phys.
  Rev. X}\ }\textbf {\bibinfo {volume} {8}},\ \bibinfo {pages} {011050}
  (\bibinfo {year} {2018})}\BibitemShut {NoStop}%
\bibitem [{\citenamefont {Ortiz-Ambriz}\ and\ \citenamefont
  {Tierno}(2016)}]{OrtizAmbriz16}%
  \BibitemOpen
  \bibfield  {author} {\bibinfo {author} {\bibfnamefont {A.}~\bibnamefont
  {Ortiz-Ambriz}}\ and\ \bibinfo {author} {\bibfnamefont {P.}~\bibnamefont
  {Tierno}},\ }\bibfield  {title} {\enquote {\bibinfo {title} {Engineering of
  frustration in colloidal artificial ices realized on microfeatured grooved
  lattices},}\ }\href {\doibase 10.1038/ncomms10575} {\bibfield  {journal}
  {\bibinfo  {journal} {Nature Commun.}\ }\textbf {\bibinfo {volume} {7}},\
  \bibinfo {pages} {10575} (\bibinfo {year} {2016})}\BibitemShut {NoStop}%
\bibitem [{\citenamefont {Stoop}\ \emph {et~al.}(2020)\citenamefont {Stoop},
  \citenamefont {Straube}, \citenamefont {Johansen},\ and\ \citenamefont
  {Tierno}}]{Stoop20}%
  \BibitemOpen
  \bibfield  {author} {\bibinfo {author} {\bibfnamefont {R.~L.}\ \bibnamefont
  {Stoop}}, \bibinfo {author} {\bibfnamefont {A.~V.}\ \bibnamefont {Straube}},
  \bibinfo {author} {\bibfnamefont {T.~H.}\ \bibnamefont {Johansen}}, \ and\
  \bibinfo {author} {\bibfnamefont {P.}~\bibnamefont {Tierno}},\ }\bibfield
  {title} {\enquote {\bibinfo {title} {Collective directional locking of
  colloidal monolayers on a periodic substrate},}\ }\href {\doibase
  10.1103/PhysRevLett.124.058002} {\bibfield  {journal} {\bibinfo  {journal}
  {Phys. Rev. Lett.}\ }\textbf {\bibinfo {volume} {124}},\ \bibinfo {pages}
  {058002} (\bibinfo {year} {2020})}\BibitemShut {NoStop}%
\bibitem [{\citenamefont {Reichhardt}\ and\ \citenamefont
  {Reichhardt}(2021{\natexlab{a}})}]{Reichhardt21}%
  \BibitemOpen
  \bibfield  {author} {\bibinfo {author} {\bibfnamefont {C.}~\bibnamefont
  {Reichhardt}}\ and\ \bibinfo {author} {\bibfnamefont {C.~J.~O.}\ \bibnamefont
  {Reichhardt}},\ }\bibfield  {title} {\enquote {\bibinfo {title} {Active
  matter commensuration and frustration effects on periodic substrates},}\
  }\href {\doibase 10.1103/PhysRevE.103.022602} {\bibfield  {journal} {\bibinfo
   {journal} {Phys. Rev. E}\ }\textbf {\bibinfo {volume} {103}},\ \bibinfo
  {pages} {022602} (\bibinfo {year} {2021}{\natexlab{a}})}\BibitemShut
  {NoStop}%
\bibitem [{\citenamefont {Lewenstein}\ \emph {et~al.}(2007)\citenamefont
  {Lewenstein}, \citenamefont {Sanpera}, \citenamefont {Ahufinger},
  \citenamefont {Damski}, \citenamefont {Sen(De)},\ and\ \citenamefont
  {Sen}}]{Lewenstein07}%
  \BibitemOpen
  \bibfield  {author} {\bibinfo {author} {\bibfnamefont {M.}~\bibnamefont
  {Lewenstein}}, \bibinfo {author} {\bibfnamefont {A.}~\bibnamefont {Sanpera}},
  \bibinfo {author} {\bibfnamefont {V.}~\bibnamefont {Ahufinger}}, \bibinfo
  {author} {\bibfnamefont {B.}~\bibnamefont {Damski}}, \bibinfo {author}
  {\bibfnamefont {A.}~\bibnamefont {Sen(De)}}, \ and\ \bibinfo {author}
  {\bibfnamefont {U.}~\bibnamefont {Sen}},\ }\bibfield  {title} {\enquote
  {\bibinfo {title} {Ultracold atomic gases in optical lattices: mimicking
  condensed matter physics and beyond},}\ }\href {\doibase
  10.1080/00018730701223200} {\bibfield  {journal} {\bibinfo  {journal} {Adv.
  Phys.}\ }\textbf {\bibinfo {volume} {56}},\ \bibinfo {pages} {243--379}
  (\bibinfo {year} {2007})}\BibitemShut {NoStop}%
\bibitem [{\citenamefont {Huang}\ \emph {et~al.}(2021)\citenamefont {Huang},
  \citenamefont {Li}, \citenamefont {Reichhardt}, \citenamefont {Reichhardt},\
  and\ \citenamefont {Feng}}]{Huang21}%
  \BibitemOpen
  \bibfield  {author} {\bibinfo {author} {\bibfnamefont {Y.}~\bibnamefont
  {Huang}}, \bibinfo {author} {\bibfnamefont {W.}~\bibnamefont {Li}}, \bibinfo
  {author} {\bibfnamefont {C.}~\bibnamefont {Reichhardt}}, \bibinfo {author}
  {\bibfnamefont {C.~J.~O.}\ \bibnamefont {Reichhardt}}, \ and\ \bibinfo
  {author} {\bibfnamefont {Y.}~\bibnamefont {Feng}},\ }\bibfield  {title}
  {\enquote {\bibinfo {title} {Phonon spectra of a two-dimensional solid dusty
  plasma modified by two-dimensional periodic substrates},}\ }\href@noop {}
  {\bibfield  {journal} {\bibinfo  {journal} {arXiv:2109.11355}\ } (\bibinfo
  {year} {2021})}\BibitemShut {NoStop}%
\bibitem [{\citenamefont {Reichhardt}\ \emph {et~al.}(2015)\citenamefont
  {Reichhardt}, \citenamefont {Ray},\ and\ \citenamefont
  {Reichhardt}}]{Reichhardt15a}%
  \BibitemOpen
  \bibfield  {author} {\bibinfo {author} {\bibfnamefont {C.}~\bibnamefont
  {Reichhardt}}, \bibinfo {author} {\bibfnamefont {D.}~\bibnamefont {Ray}}, \
  and\ \bibinfo {author} {\bibfnamefont {C.~J.~Olson}\ \bibnamefont
  {Reichhardt}},\ }\bibfield  {title} {\enquote {\bibinfo {title} {Quantized
  transport for a skyrmion moving on a two-dimensional periodic substrate},}\
  }\href {\doibase 10.1103/PhysRevB.91.104426} {\bibfield  {journal} {\bibinfo
  {journal} {Phys. Rev. B}\ }\textbf {\bibinfo {volume} {91}},\ \bibinfo
  {pages} {104426} (\bibinfo {year} {2015})}\BibitemShut {NoStop}%
\bibitem [{\citenamefont {Feilhauer}\ \emph {et~al.}(2020)\citenamefont
  {Feilhauer}, \citenamefont {Saha}, \citenamefont {Tobik}, \citenamefont
  {Zelent}, \citenamefont {Heyderman},\ and\ \citenamefont
  {Mruczkiewicz}}]{Feilhauer20}%
  \BibitemOpen
  \bibfield  {author} {\bibinfo {author} {\bibfnamefont {J.}~\bibnamefont
  {Feilhauer}}, \bibinfo {author} {\bibfnamefont {S.}~\bibnamefont {Saha}},
  \bibinfo {author} {\bibfnamefont {J.}~\bibnamefont {Tobik}}, \bibinfo
  {author} {\bibfnamefont {M.}~\bibnamefont {Zelent}}, \bibinfo {author}
  {\bibfnamefont {L.~J.}\ \bibnamefont {Heyderman}}, \ and\ \bibinfo {author}
  {\bibfnamefont {M.}~\bibnamefont {Mruczkiewicz}},\ }\bibfield  {title}
  {\enquote {\bibinfo {title} {Controlled motion of skyrmions in a magnetic
  antidot lattice},}\ }\href {\doibase 10.1103/PhysRevB.102.184425} {\bibfield
  {journal} {\bibinfo  {journal} {Phys. Rev. B}\ }\textbf {\bibinfo {volume}
  {102}},\ \bibinfo {pages} {184425} (\bibinfo {year} {2020})}\BibitemShut
  {NoStop}%
\bibitem [{\citenamefont {Nisoli}\ \emph {et~al.}(2013)\citenamefont {Nisoli},
  \citenamefont {Moessner},\ and\ \citenamefont {Schiffer}}]{Nisoli13}%
  \BibitemOpen
  \bibfield  {author} {\bibinfo {author} {\bibfnamefont {C.}~\bibnamefont
  {Nisoli}}, \bibinfo {author} {\bibfnamefont {R.}~\bibnamefont {Moessner}}, \
  and\ \bibinfo {author} {\bibfnamefont {P.}~\bibnamefont {Schiffer}},\
  }\bibfield  {title} {\enquote {\bibinfo {title} {Colloquium: Artificial spin
  ice: Designing and imaging magnetic frustration},}\ }\href {\doibase
  10.1103/RevModPhys.85.1473} {\bibfield  {journal} {\bibinfo  {journal} {Rev.
  Mod. Phys.}\ }\textbf {\bibinfo {volume} {85}},\ \bibinfo {pages}
  {1473--1490} (\bibinfo {year} {2013})}\BibitemShut {NoStop}%
\bibitem [{\citenamefont {Hasnain}\ \emph {et~al.}(2014)\citenamefont
  {Hasnain}, \citenamefont {Jungblut}, \citenamefont {Tr{\" o}ster},\ and\
  \citenamefont {Dellago}}]{Hasnain14}%
  \BibitemOpen
  \bibfield  {author} {\bibinfo {author} {\bibfnamefont {J.}~\bibnamefont
  {Hasnain}}, \bibinfo {author} {\bibfnamefont {S.}~\bibnamefont {Jungblut}},
  \bibinfo {author} {\bibfnamefont {A.}~\bibnamefont {Tr{\" o}ster}}, \ and\
  \bibinfo {author} {\bibfnamefont {C.}~\bibnamefont {Dellago}},\ }\bibfield
  {title} {\enquote {\bibinfo {title} {Frictional dynamics of stiff monolayers:
  from nucleation dynamics to thermal sliding},}\ }\href {\doibase
  10.1039/C4NR01790K} {\bibfield  {journal} {\bibinfo  {journal} {Nanoscale}\
  }\textbf {\bibinfo {volume} {6}},\ \bibinfo {pages} {10161--10168} (\bibinfo
  {year} {2014})}\BibitemShut {NoStop}%
\bibitem [{\citenamefont {Koplik}\ and\ \citenamefont
  {Drazer}(2010)}]{Koplik10}%
  \BibitemOpen
  \bibfield  {author} {\bibinfo {author} {\bibfnamefont {J.}~\bibnamefont
  {Koplik}}\ and\ \bibinfo {author} {\bibfnamefont {G.}~\bibnamefont
  {Drazer}},\ }\bibfield  {title} {\enquote {\bibinfo {title} {Nanoscale
  simulations of directional locking},}\ }\href {\doibase 10.1063/1.3429297}
  {\bibfield  {journal} {\bibinfo  {journal} {Phys. Fluids}\ }\textbf {\bibinfo
  {volume} {22}},\ \bibinfo {pages} {052005} (\bibinfo {year}
  {2010})}\BibitemShut {NoStop}%
\bibitem [{\citenamefont {\c{C}am}\ \emph {et~al.}(2021)\citenamefont
  {\c{C}am}, \citenamefont {Lichter},\ and\ \citenamefont {Goedde}}]{Cam21}%
  \BibitemOpen
  \bibfield  {author} {\bibinfo {author} {\bibfnamefont {M.}~\bibnamefont
  {\c{C}am}}, \bibinfo {author} {\bibfnamefont {S.}~\bibnamefont {Lichter}}, \
  and\ \bibinfo {author} {\bibfnamefont {C.~G.}\ \bibnamefont {Goedde}},\
  }\bibfield  {title} {\enquote {\bibinfo {title} {Kink propagation and solute
  partitioning in an atomic monolayer on a substrate},}\ }\href {\doibase
  10.1103/PhysRevE.104.L022801} {\bibfield  {journal} {\bibinfo  {journal}
  {Phys. Rev. E}\ }\textbf {\bibinfo {volume} {104}},\ \bibinfo {pages}
  {L022801} (\bibinfo {year} {2021})}\BibitemShut {NoStop}%
\bibitem [{\citenamefont {McDermott}\ \emph
  {et~al.}(2013{\natexlab{b}})\citenamefont {McDermott}, \citenamefont
  {Amelang}, \citenamefont {Reichhardt},\ and\ \citenamefont
  {Reichhardt}}]{McDermott13a}%
  \BibitemOpen
  \bibfield  {author} {\bibinfo {author} {\bibfnamefont {D.}~\bibnamefont
  {McDermott}}, \bibinfo {author} {\bibfnamefont {J.}~\bibnamefont {Amelang}},
  \bibinfo {author} {\bibfnamefont {C.~J.~Olson}\ \bibnamefont {Reichhardt}}, \
  and\ \bibinfo {author} {\bibfnamefont {C.}~\bibnamefont {Reichhardt}},\
  }\bibfield  {title} {\enquote {\bibinfo {title} {Dynamic regimes for driven
  colloidal particles on a periodic substrate at commensurate and
  incommensurate fillings},}\ }\href {\doibase 10.1103/PhysRevE.88.062301}
  {\bibfield  {journal} {\bibinfo  {journal} {Phys. Rev. E}\ }\textbf {\bibinfo
  {volume} {88}},\ \bibinfo {pages} {062301} (\bibinfo {year}
  {2013}{\natexlab{b}})}\BibitemShut {NoStop}%
\bibitem [{\citenamefont {Reichhardt}\ \emph
  {et~al.}(1998{\natexlab{b}})\citenamefont {Reichhardt}, \citenamefont
  {Olson},\ and\ \citenamefont {Nori}}]{Reichhardt98}%
  \BibitemOpen
  \bibfield  {author} {\bibinfo {author} {\bibfnamefont {C.}~\bibnamefont
  {Reichhardt}}, \bibinfo {author} {\bibfnamefont {C.~J.}\ \bibnamefont
  {Olson}}, \ and\ \bibinfo {author} {\bibfnamefont {F.}~\bibnamefont {Nori}},\
  }\bibfield  {title} {\enquote {\bibinfo {title} {Nonequilibrium dynamic
  phases and plastic flow of driven vortex lattices in superconductors with
  periodic arrays of pinning sites},}\ }\href {\doibase
  10.1103/PhysRevB.58.6534} {\bibfield  {journal} {\bibinfo  {journal} {Phys.
  Rev. B}\ }\textbf {\bibinfo {volume} {58}},\ \bibinfo {pages} {6534--6564}
  (\bibinfo {year} {1998}{\natexlab{b}})}\BibitemShut {NoStop}%
\bibitem [{\citenamefont {Durkin}\ \emph {et~al.}(2016)\citenamefont {Durkin},
  \citenamefont {Mondragon-Shem}, \citenamefont {Eley}, \citenamefont
  {Hughes},\ and\ \citenamefont {Mason}}]{Durkin16}%
  \BibitemOpen
  \bibfield  {author} {\bibinfo {author} {\bibfnamefont {M.}~\bibnamefont
  {Durkin}}, \bibinfo {author} {\bibfnamefont {I.}~\bibnamefont
  {Mondragon-Shem}}, \bibinfo {author} {\bibfnamefont {S.}~\bibnamefont
  {Eley}}, \bibinfo {author} {\bibfnamefont {T.~L.}\ \bibnamefont {Hughes}}, \
  and\ \bibinfo {author} {\bibfnamefont {N.}~\bibnamefont {Mason}},\ }\bibfield
   {title} {\enquote {\bibinfo {title} {History-dependent dissipative vortex
  dynamics in superconducting arrays},}\ }\href {\doibase
  10.1103/PhysRevB.94.024510} {\bibfield  {journal} {\bibinfo  {journal} {Phys.
  Rev. B}\ }\textbf {\bibinfo {volume} {94}},\ \bibinfo {pages} {024510}
  (\bibinfo {year} {2016})}\BibitemShut {NoStop}%
\bibitem [{\citenamefont {Hastings}\ \emph {et~al.}(2003)\citenamefont
  {Hastings}, \citenamefont {Olson~Reichhardt},\ and\ \citenamefont
  {Reichhardt}}]{Hastings03}%
  \BibitemOpen
  \bibfield  {author} {\bibinfo {author} {\bibfnamefont {M.~B.}\ \bibnamefont
  {Hastings}}, \bibinfo {author} {\bibfnamefont {C.~J.}\ \bibnamefont
  {Olson~Reichhardt}}, \ and\ \bibinfo {author} {\bibfnamefont
  {C.}~\bibnamefont {Reichhardt}},\ }\bibfield  {title} {\enquote {\bibinfo
  {title} {Depinning by fracture in a glassy background},}\ }\href {\doibase
  10.1103/PhysRevLett.90.098302} {\bibfield  {journal} {\bibinfo  {journal}
  {Phys. Rev. Lett.}\ }\textbf {\bibinfo {volume} {90}},\ \bibinfo {pages}
  {098302} (\bibinfo {year} {2003})}\BibitemShut {NoStop}%
\bibitem [{\citenamefont {Habdas}\ \emph {et~al.}(2004)\citenamefont {Habdas},
  \citenamefont {Schaar}, \citenamefont {Levitt},\ and\ \citenamefont
  {Weeks}}]{Habdas04}%
  \BibitemOpen
  \bibfield  {author} {\bibinfo {author} {\bibfnamefont {P.}~\bibnamefont
  {Habdas}}, \bibinfo {author} {\bibfnamefont {D.}~\bibnamefont {Schaar}},
  \bibinfo {author} {\bibfnamefont {A.~C.}\ \bibnamefont {Levitt}}, \ and\
  \bibinfo {author} {\bibfnamefont {E.~R.}\ \bibnamefont {Weeks}},\ }\bibfield
  {title} {\enquote {\bibinfo {title} {Forced motion of a probe particle near
  the colloidal glass transition},}\ }\href {\doibase
  10.1209/epl/i2004-10075-y} {\bibfield  {journal} {\bibinfo  {journal}
  {Europhys. Lett.}\ }\textbf {\bibinfo {volume} {67}},\ \bibinfo {pages}
  {477--483} (\bibinfo {year} {2004})}\BibitemShut {NoStop}%
\bibitem [{\citenamefont {Squires}\ and\ \citenamefont
  {Brady}(2005)}]{Squires05}%
  \BibitemOpen
  \bibfield  {author} {\bibinfo {author} {\bibfnamefont {T.~M.}\ \bibnamefont
  {Squires}}\ and\ \bibinfo {author} {\bibfnamefont {J.~F.}\ \bibnamefont
  {Brady}},\ }\bibfield  {title} {\enquote {\bibinfo {title} {A simple paradigm
  for active and nonlinear microrheology},}\ }\href {\doibase
  10.1063/1.1960607} {\bibfield  {journal} {\bibinfo  {journal} {Phys. Fluids}\
  }\textbf {\bibinfo {volume} {17}},\ \bibinfo {pages} {073101} (\bibinfo
  {year} {2005})}\BibitemShut {NoStop}%
\bibitem [{\citenamefont {Voigtmann}\ and\ \citenamefont
  {Fuchs}(2013)}]{Voigtmann13}%
  \BibitemOpen
  \bibfield  {author} {\bibinfo {author} {\bibfnamefont {Th.}\ \bibnamefont
  {Voigtmann}}\ and\ \bibinfo {author} {\bibfnamefont {M.}~\bibnamefont
  {Fuchs}},\ }\bibfield  {title} {\enquote {\bibinfo {title} {Force-driven
  micro-rheology},}\ }\href {\doibase 10.1140/epjst/e2013-02060-5} {\bibfield
  {journal} {\bibinfo  {journal} {Eur. Phys. J. Spec. Top.}\ }\textbf {\bibinfo
  {volume} {222}},\ \bibinfo {pages} {2819--2833} (\bibinfo {year}
  {2013})}\BibitemShut {NoStop}%
\bibitem [{\citenamefont {Zia}(2018)}]{Zia18}%
  \BibitemOpen
  \bibfield  {author} {\bibinfo {author} {\bibfnamefont {R.~N.}\ \bibnamefont
  {Zia}},\ }\bibfield  {title} {\enquote {\bibinfo {title} {Active and passive
  microrheology: Theory and simulation},}\ }\href {\doibase
  10.1146/annurev-fluid-122316-044514} {\bibfield  {journal} {\bibinfo
  {journal} {Ann. Rev. Fluid Mech.}\ }\textbf {\bibinfo {volume} {50}},\
  \bibinfo {pages} {371} (\bibinfo {year} {2018})}\BibitemShut {NoStop}%
\bibitem [{\citenamefont {Gazuz}\ \emph {et~al.}(2009)\citenamefont {Gazuz},
  \citenamefont {Puertas}, \citenamefont {Voigtmann},\ and\ \citenamefont
  {Fuchs}}]{Gazuz09}%
  \BibitemOpen
  \bibfield  {author} {\bibinfo {author} {\bibfnamefont {I.}~\bibnamefont
  {Gazuz}}, \bibinfo {author} {\bibfnamefont {A.~M.}\ \bibnamefont {Puertas}},
  \bibinfo {author} {\bibfnamefont {Th.}\ \bibnamefont {Voigtmann}}, \ and\
  \bibinfo {author} {\bibfnamefont {M.}~\bibnamefont {Fuchs}},\ }\bibfield
  {title} {\enquote {\bibinfo {title} {Active and nonlinear microrheology in
  dense colloidal suspensions},}\ }\href {\doibase
  10.1103/PhysRevLett.102.248302} {\bibfield  {journal} {\bibinfo  {journal}
  {Phys. Rev. Lett.}\ }\textbf {\bibinfo {volume} {102}},\ \bibinfo {pages}
  {248302} (\bibinfo {year} {2009})}\BibitemShut {NoStop}%
\bibitem [{\citenamefont {Reichhardt}\ and\ \citenamefont
  {Reichhardt}(2021{\natexlab{b}})}]{Reichhardt21a}%
  \BibitemOpen
  \bibfield  {author} {\bibinfo {author} {\bibfnamefont {C.}~\bibnamefont
  {Reichhardt}}\ and\ \bibinfo {author} {\bibfnamefont {C.~J.~O.}\ \bibnamefont
  {Reichhardt}},\ }\bibfield  {title} {\enquote {\bibinfo {title} {Dynamics and
  nonmonotonic drag for individually driven skyrmions},}\ }\href {\doibase
  10.1103/PhysRevB.104.064441} {\bibfield  {journal} {\bibinfo  {journal}
  {Phys. Rev. B}\ }\textbf {\bibinfo {volume} {104}},\ \bibinfo {pages}
  {064441} (\bibinfo {year} {2021}{\natexlab{b}})}\BibitemShut {NoStop}%
\bibitem [{\citenamefont {Winter}\ \emph {et~al.}(2012)\citenamefont {Winter},
  \citenamefont {Horbach}, \citenamefont {Virnau},\ and\ \citenamefont
  {Binder}}]{Winter12}%
  \BibitemOpen
  \bibfield  {author} {\bibinfo {author} {\bibfnamefont {D.}~\bibnamefont
  {Winter}}, \bibinfo {author} {\bibfnamefont {J.}~\bibnamefont {Horbach}},
  \bibinfo {author} {\bibfnamefont {P.}~\bibnamefont {Virnau}}, \ and\ \bibinfo
  {author} {\bibfnamefont {K.}~\bibnamefont {Binder}},\ }\bibfield  {title}
  {\enquote {\bibinfo {title} {Active nonlinear microrheology in a
  glass-forming {Y}ukawa fluid},}\ }\href {\doibase
  10.1103/PhysRevLett.108.028303} {\bibfield  {journal} {\bibinfo  {journal}
  {Phys. Rev. Lett.}\ }\textbf {\bibinfo {volume} {108}},\ \bibinfo {pages}
  {028303} (\bibinfo {year} {2012})}\BibitemShut {NoStop}%
\bibitem [{\citenamefont {\c{S}enbil}\ \emph {et~al.}(2019)\citenamefont
  {\c{S}enbil}, \citenamefont {Gruber}, \citenamefont {Zhang}, \citenamefont
  {Fuchs},\ and\ \citenamefont {Scheffold}}]{Senbil19}%
  \BibitemOpen
  \bibfield  {author} {\bibinfo {author} {\bibfnamefont {N.}~\bibnamefont
  {\c{S}enbil}}, \bibinfo {author} {\bibfnamefont {M.}~\bibnamefont {Gruber}},
  \bibinfo {author} {\bibfnamefont {C.}~\bibnamefont {Zhang}}, \bibinfo
  {author} {\bibfnamefont {M.}~\bibnamefont {Fuchs}}, \ and\ \bibinfo {author}
  {\bibfnamefont {F.}~\bibnamefont {Scheffold}},\ }\bibfield  {title} {\enquote
  {\bibinfo {title} {Observation of strongly heterogeneous dynamics at the
  depinning transition in a colloidal glass},}\ }\href {\doibase
  10.1103/PhysRevLett.122.108002} {\bibfield  {journal} {\bibinfo  {journal}
  {Phys. Rev. Lett.}\ }\textbf {\bibinfo {volume} {122}},\ \bibinfo {pages}
  {108002} (\bibinfo {year} {2019})}\BibitemShut {NoStop}%
\bibitem [{\citenamefont {Yu}\ \emph {et~al.}(2020)\citenamefont {Yu},
  \citenamefont {Rahbari}, \citenamefont {Kawasaki}, \citenamefont {Park},\
  and\ \citenamefont {Lee}}]{Yu20}%
  \BibitemOpen
  \bibfield  {author} {\bibinfo {author} {\bibfnamefont {J.~W.}\ \bibnamefont
  {Yu}}, \bibinfo {author} {\bibfnamefont {S.~H.~E.}\ \bibnamefont {Rahbari}},
  \bibinfo {author} {\bibfnamefont {T.}~\bibnamefont {Kawasaki}}, \bibinfo
  {author} {\bibfnamefont {H.}~\bibnamefont {Park}}, \ and\ \bibinfo {author}
  {\bibfnamefont {W.~B.}\ \bibnamefont {Lee}},\ }\bibfield  {title} {\enquote
  {\bibinfo {title} {Active microrheology of a bulk metallic glass},}\ }\href
  {\doibase 10.1126/sciadv.aba8766} {\bibfield  {journal} {\bibinfo  {journal}
  {Sci. Adv.}\ }\textbf {\bibinfo {volume} {6}} (\bibinfo {year} {2020}),\
  10.1126/sciadv.aba8766}\BibitemShut {NoStop}%
\bibitem [{\citenamefont {Drocco}\ \emph {et~al.}(2005)\citenamefont {Drocco},
  \citenamefont {Hastings}, \citenamefont {Reichhardt},\ and\ \citenamefont
  {Reichhardt}}]{Drocco05}%
  \BibitemOpen
  \bibfield  {author} {\bibinfo {author} {\bibfnamefont {J.~A.}\ \bibnamefont
  {Drocco}}, \bibinfo {author} {\bibfnamefont {M.~B.}\ \bibnamefont
  {Hastings}}, \bibinfo {author} {\bibfnamefont {C.~J.~Olson}\ \bibnamefont
  {Reichhardt}}, \ and\ \bibinfo {author} {\bibfnamefont {C.}~\bibnamefont
  {Reichhardt}},\ }\bibfield  {title} {\enquote {\bibinfo {title} {Multiscaling
  at point {$J$}: Jamming is a critical phenomenon},}\ }\href {\doibase
  10.1103/PhysRevLett.95.088001} {\bibfield  {journal} {\bibinfo  {journal}
  {Phys. Rev. Lett.}\ }\textbf {\bibinfo {volume} {95}},\ \bibinfo {pages}
  {088001} (\bibinfo {year} {2005})}\BibitemShut {NoStop}%
\bibitem [{\citenamefont {Candelier}\ and\ \citenamefont
  {Dauchot}(2010)}]{Candelier10}%
  \BibitemOpen
  \bibfield  {author} {\bibinfo {author} {\bibfnamefont {R.}~\bibnamefont
  {Candelier}}\ and\ \bibinfo {author} {\bibfnamefont {O.}~\bibnamefont
  {Dauchot}},\ }\bibfield  {title} {\enquote {\bibinfo {title} {Journey of an
  intruder through the fluidization and jamming transitions of a dense granular
  media},}\ }\href {\doibase 10.1103/PhysRevE.81.011304} {\bibfield  {journal}
  {\bibinfo  {journal} {Phys. Rev. E}\ }\textbf {\bibinfo {volume} {81}},\
  \bibinfo {pages} {011304} (\bibinfo {year} {2010})}\BibitemShut {NoStop}%
\bibitem [{\citenamefont {Kolb}\ \emph {et~al.}(2013)\citenamefont {Kolb},
  \citenamefont {Cixous}, \citenamefont {Gaudouen},\ and\ \citenamefont
  {Darnige}}]{Kolb13}%
  \BibitemOpen
  \bibfield  {author} {\bibinfo {author} {\bibfnamefont {E.}~\bibnamefont
  {Kolb}}, \bibinfo {author} {\bibfnamefont {P.}~\bibnamefont {Cixous}},
  \bibinfo {author} {\bibfnamefont {N.}~\bibnamefont {Gaudouen}}, \ and\
  \bibinfo {author} {\bibfnamefont {T.}~\bibnamefont {Darnige}},\ }\bibfield
  {title} {\enquote {\bibinfo {title} {Rigid intruder inside a two-dimensional
  dense granular flow: Drag force and cavity formation},}\ }\href {\doibase
  10.1103/PhysRevE.87.032207} {\bibfield  {journal} {\bibinfo  {journal} {Phys.
  Rev. E}\ }\textbf {\bibinfo {volume} {87}},\ \bibinfo {pages} {032207}
  (\bibinfo {year} {2013})}\BibitemShut {NoStop}%
\bibitem [{\citenamefont {Dullens}\ and\ \citenamefont
  {Bechinger}(2011)}]{Dullens11}%
  \BibitemOpen
  \bibfield  {author} {\bibinfo {author} {\bibfnamefont {R.~P.~A.}\
  \bibnamefont {Dullens}}\ and\ \bibinfo {author} {\bibfnamefont
  {C.}~\bibnamefont {Bechinger}},\ }\bibfield  {title} {\enquote {\bibinfo
  {title} {Shear thinning and local melting of colloidal crystals},}\ }\href
  {\doibase 10.1103/PhysRevLett.107.138301} {\bibfield  {journal} {\bibinfo
  {journal} {Phys. Rev. Lett.}\ }\textbf {\bibinfo {volume} {107}},\ \bibinfo
  {pages} {138301} (\bibinfo {year} {2011})}\BibitemShut {NoStop}%
\bibitem [{\citenamefont {Abaurrea-Velasco}\ \emph {et~al.}(2020)\citenamefont
  {Abaurrea-Velasco}, \citenamefont {Lozano}, \citenamefont {Bechinger},\ and\
  \citenamefont {de~Graaf}}]{AbaurreaVelasco20}%
  \BibitemOpen
  \bibfield  {author} {\bibinfo {author} {\bibfnamefont {C.}~\bibnamefont
  {Abaurrea-Velasco}}, \bibinfo {author} {\bibfnamefont {C.}~\bibnamefont
  {Lozano}}, \bibinfo {author} {\bibfnamefont {C.}~\bibnamefont {Bechinger}}, \
  and\ \bibinfo {author} {\bibfnamefont {J.}~\bibnamefont {de~Graaf}},\
  }\bibfield  {title} {\enquote {\bibinfo {title} {Autonomously probing
  viscoelasticity in disordered suspensions},}\ }\href {\doibase
  10.1103/PhysRevLett.125.258002} {\bibfield  {journal} {\bibinfo  {journal}
  {Phys. Rev. Lett.}\ }\textbf {\bibinfo {volume} {125}},\ \bibinfo {pages}
  {258002} (\bibinfo {year} {2020})}\BibitemShut {NoStop}%
\bibitem [{\citenamefont {Straver}\ \emph {et~al.}(2008)\citenamefont
  {Straver}, \citenamefont {Hoffman}, \citenamefont {Auslaender}, \citenamefont
  {Rugar},\ and\ \citenamefont {Moler}}]{Straver08}%
  \BibitemOpen
  \bibfield  {author} {\bibinfo {author} {\bibfnamefont {E.~W.~J.}\
  \bibnamefont {Straver}}, \bibinfo {author} {\bibfnamefont {J.~E.}\
  \bibnamefont {Hoffman}}, \bibinfo {author} {\bibfnamefont {O.~M.}\
  \bibnamefont {Auslaender}}, \bibinfo {author} {\bibfnamefont
  {D.}~\bibnamefont {Rugar}}, \ and\ \bibinfo {author} {\bibfnamefont {K.~A.}\
  \bibnamefont {Moler}},\ }\bibfield  {title} {\enquote {\bibinfo {title}
  {Controlled manipulation of individual vortices in a superconductor},}\
  }\href {\doibase 10.1063/1.3000963} {\bibfield  {journal} {\bibinfo
  {journal} {Appl. Phys. Lett.}\ }\textbf {\bibinfo {volume} {93}},\ \bibinfo
  {pages} {172514} (\bibinfo {year} {2008})}\BibitemShut {NoStop}%
\bibitem [{\citenamefont {Reichhardt}(2009)}]{Reichhardt09a}%
  \BibitemOpen
  \bibfield  {author} {\bibinfo {author} {\bibfnamefont {C.}~\bibnamefont
  {Reichhardt}},\ }\bibfield  {title} {\enquote {\bibinfo {title} {Vortices
  wiggled and dragged},}\ }\href {\doibase 10.1038/nphys1169} {\bibfield
  {journal} {\bibinfo  {journal} {Nature Phys.}\ }\textbf {\bibinfo {volume}
  {5}},\ \bibinfo {pages} {15--16} (\bibinfo {year} {2009})}\BibitemShut
  {NoStop}%
\bibitem [{\citenamefont {Auslaender}\ \emph {et~al.}(2009)\citenamefont
  {Auslaender}, \citenamefont {Luan}, \citenamefont {Straver}, \citenamefont
  {Hoffman}, \citenamefont {Koshnick}, \citenamefont {Zeldov}, \citenamefont
  {Bonn}, \citenamefont {Liang}, \citenamefont {Hardy},\ and\ \citenamefont
  {Moler}}]{Auslaender09}%
  \BibitemOpen
  \bibfield  {author} {\bibinfo {author} {\bibfnamefont {O.~M.}\ \bibnamefont
  {Auslaender}}, \bibinfo {author} {\bibfnamefont {L.}~\bibnamefont {Luan}},
  \bibinfo {author} {\bibfnamefont {E.~W.~J.}\ \bibnamefont {Straver}},
  \bibinfo {author} {\bibfnamefont {J.~E.}\ \bibnamefont {Hoffman}}, \bibinfo
  {author} {\bibfnamefont {N.~C.}\ \bibnamefont {Koshnick}}, \bibinfo {author}
  {\bibfnamefont {E.}~\bibnamefont {Zeldov}}, \bibinfo {author} {\bibfnamefont
  {D.~A.}\ \bibnamefont {Bonn}}, \bibinfo {author} {\bibfnamefont
  {R.}~\bibnamefont {Liang}}, \bibinfo {author} {\bibfnamefont {W.~N.}\
  \bibnamefont {Hardy}}, \ and\ \bibinfo {author} {\bibfnamefont {K.~A.}\
  \bibnamefont {Moler}},\ }\bibfield  {title} {\enquote {\bibinfo {title}
  {Mechanics of individual isolated vortices in a cuprate superconductor},}\
  }\href {\doibase 10.1038/NPHYS1127} {\bibfield  {journal} {\bibinfo
  {journal} {Nature Phys.}\ }\textbf {\bibinfo {volume} {5}},\ \bibinfo {pages}
  {35--39} (\bibinfo {year} {2009})}\BibitemShut {NoStop}%
\bibitem [{\citenamefont {Hanneken}\ \emph {et~al.}(2016)\citenamefont
  {Hanneken}, \citenamefont {Kubetzka}, \citenamefont {von Bergmann},\ and\
  \citenamefont {Wiesendanger}}]{Hanneken16}%
  \BibitemOpen
  \bibfield  {author} {\bibinfo {author} {\bibfnamefont {C.}~\bibnamefont
  {Hanneken}}, \bibinfo {author} {\bibfnamefont {A.}~\bibnamefont {Kubetzka}},
  \bibinfo {author} {\bibfnamefont {K.}~\bibnamefont {von Bergmann}}, \ and\
  \bibinfo {author} {\bibfnamefont {R.}~\bibnamefont {Wiesendanger}},\
  }\bibfield  {title} {\enquote {\bibinfo {title} {Pinning and movement of
  individual nanoscale magnetic skyrmions via defects},}\ }\href {\doibase
  10.1088/1367-2630/18/5/055009} {\bibfield  {journal} {\bibinfo  {journal}
  {New J. Phys.}\ }\textbf {\bibinfo {volume} {18}},\ \bibinfo {pages} {055009}
  (\bibinfo {year} {2016})}\BibitemShut {NoStop}%
\bibitem [{\citenamefont {Casiraghi}\ \emph {et~al.}(2019)\citenamefont
  {Casiraghi}, \citenamefont {Corte-Le{\' o}n}, \citenamefont {Vafaee},
  \citenamefont {Garcia-Sanchez}, \citenamefont {Durin}, \citenamefont
  {Pasquale}, \citenamefont {Jakob}, \citenamefont {Kl{\" a}ui},\ and\
  \citenamefont {Kazakova}}]{Casiraghi19}%
  \BibitemOpen
  \bibfield  {author} {\bibinfo {author} {\bibfnamefont {A.}~\bibnamefont
  {Casiraghi}}, \bibinfo {author} {\bibfnamefont {H.}~\bibnamefont {Corte-Le{\'
  o}n}}, \bibinfo {author} {\bibfnamefont {M.}~\bibnamefont {Vafaee}}, \bibinfo
  {author} {\bibfnamefont {F.}~\bibnamefont {Garcia-Sanchez}}, \bibinfo
  {author} {\bibfnamefont {G.}~\bibnamefont {Durin}}, \bibinfo {author}
  {\bibfnamefont {M.}~\bibnamefont {Pasquale}}, \bibinfo {author}
  {\bibfnamefont {G.}~\bibnamefont {Jakob}}, \bibinfo {author} {\bibfnamefont
  {M.}~\bibnamefont {Kl{\" a}ui}}, \ and\ \bibinfo {author} {\bibfnamefont
  {O.}~\bibnamefont {Kazakova}},\ }\bibfield  {title} {\enquote {\bibinfo
  {title} {Individual skyrmion manipulation by local magnetic field
  gradients},}\ }\href {\doibase 10.1038/s42005-019-0242-5} {\bibfield
  {journal} {\bibinfo  {journal} {Commun. Phys.}\ }\textbf {\bibinfo {volume}
  {2}},\ \bibinfo {pages} {145} (\bibinfo {year} {2019})}\BibitemShut {NoStop}%
\bibitem [{\citenamefont {Shapira}\ \emph {et~al.}(2015)\citenamefont
  {Shapira}, \citenamefont {Lamhot}, \citenamefont {Shpielberg}, \citenamefont
  {Kafri}, \citenamefont {Ramshaw}, \citenamefont {Bonn}, \citenamefont
  {Liang}, \citenamefont {Hardy},\ and\ \citenamefont
  {Auslaender}}]{Shapira15}%
  \BibitemOpen
  \bibfield  {author} {\bibinfo {author} {\bibfnamefont {N.}~\bibnamefont
  {Shapira}}, \bibinfo {author} {\bibfnamefont {Y.}~\bibnamefont {Lamhot}},
  \bibinfo {author} {\bibfnamefont {O.}~\bibnamefont {Shpielberg}}, \bibinfo
  {author} {\bibfnamefont {Y.}~\bibnamefont {Kafri}}, \bibinfo {author}
  {\bibfnamefont {B.~J.}\ \bibnamefont {Ramshaw}}, \bibinfo {author}
  {\bibfnamefont {D.~A.}\ \bibnamefont {Bonn}}, \bibinfo {author}
  {\bibfnamefont {R.}~\bibnamefont {Liang}}, \bibinfo {author} {\bibfnamefont
  {W.~N.}\ \bibnamefont {Hardy}}, \ and\ \bibinfo {author} {\bibfnamefont
  {O.~M.}\ \bibnamefont {Auslaender}},\ }\bibfield  {title} {\enquote {\bibinfo
  {title} {Disorder-induced power-law response of a superconducting vortex on a
  plane},}\ }\href {\doibase 10.1103/PhysRevB.92.100501} {\bibfield  {journal}
  {\bibinfo  {journal} {Phys. Rev. B}\ }\textbf {\bibinfo {volume} {92}},\
  \bibinfo {pages} {100501} (\bibinfo {year} {2015})}\BibitemShut {NoStop}%
\bibitem [{\citenamefont {Veshchunov}\ \emph {et~al.}(2016)\citenamefont
  {Veshchunov}, \citenamefont {Magrini}, \citenamefont {Mironov}, \citenamefont
  {Godin}, \citenamefont {Trebbia}, \citenamefont {Buzdin}, \citenamefont
  {Tamarat},\ and\ \citenamefont {Lounis}}]{Veshchunov16}%
  \BibitemOpen
  \bibfield  {author} {\bibinfo {author} {\bibfnamefont {I.~S.}\ \bibnamefont
  {Veshchunov}}, \bibinfo {author} {\bibfnamefont {W.}~\bibnamefont {Magrini}},
  \bibinfo {author} {\bibfnamefont {S.~V.}\ \bibnamefont {Mironov}}, \bibinfo
  {author} {\bibfnamefont {A.~G.}\ \bibnamefont {Godin}}, \bibinfo {author}
  {\bibfnamefont {J.~B.}\ \bibnamefont {Trebbia}}, \bibinfo {author}
  {\bibfnamefont {A.~I.}\ \bibnamefont {Buzdin}}, \bibinfo {author}
  {\bibfnamefont {Ph.}\ \bibnamefont {Tamarat}}, \ and\ \bibinfo {author}
  {\bibfnamefont {B.}~\bibnamefont {Lounis}},\ }\bibfield  {title} {\enquote
  {\bibinfo {title} {Optical manipulation of single flux quanta},}\ }\href
  {\doibase 10.1038/ncomms12801} {\bibfield  {journal} {\bibinfo  {journal}
  {Nature Commun.}\ }\textbf {\bibinfo {volume} {7}},\ \bibinfo {pages} {12801}
  (\bibinfo {year} {2016})}\BibitemShut {NoStop}%
\bibitem [{\citenamefont {Kremen}\ \emph {et~al.}(2016)\citenamefont {Kremen},
  \citenamefont {Wissberg}, \citenamefont {Haham}, \citenamefont {Persky},
  \citenamefont {Frenkel},\ and\ \citenamefont {Kalisky}}]{Kremen16}%
  \BibitemOpen
  \bibfield  {author} {\bibinfo {author} {\bibfnamefont {A.}~\bibnamefont
  {Kremen}}, \bibinfo {author} {\bibfnamefont {S.}~\bibnamefont {Wissberg}},
  \bibinfo {author} {\bibfnamefont {N.}~\bibnamefont {Haham}}, \bibinfo
  {author} {\bibfnamefont {E.}~\bibnamefont {Persky}}, \bibinfo {author}
  {\bibfnamefont {Y.}~\bibnamefont {Frenkel}}, \ and\ \bibinfo {author}
  {\bibfnamefont {B.}~\bibnamefont {Kalisky}},\ }\bibfield  {title} {\enquote
  {\bibinfo {title} {Mechanical control of individual superconducting
  vortices},}\ }\href {\doibase 10.1021/acs.nanolett.5b04444} {\bibfield
  {journal} {\bibinfo  {journal} {Nano Lett.}\ }\textbf {\bibinfo {volume}
  {16}},\ \bibinfo {pages} {1626--1630} (\bibinfo {year} {2016})}\BibitemShut
  {NoStop}%
\bibitem [{\citenamefont {Polshyn}\ \emph {et~al.}(2019)\citenamefont
  {Polshyn}, \citenamefont {Naibert},\ and\ \citenamefont
  {Budakian}}]{Polshyn19}%
  \BibitemOpen
  \bibfield  {author} {\bibinfo {author} {\bibfnamefont {H.}~\bibnamefont
  {Polshyn}}, \bibinfo {author} {\bibfnamefont {T.}~\bibnamefont {Naibert}}, \
  and\ \bibinfo {author} {\bibfnamefont {R.}~\bibnamefont {Budakian}},\
  }\bibfield  {title} {\enquote {\bibinfo {title} {Manipulating multivortex
  states in superconducting structures},}\ }\href {\doibase
  10.1021/acs.nanolett.9b01983} {\bibfield  {journal} {\bibinfo  {journal}
  {Nano Lett.}\ }\textbf {\bibinfo {volume} {19}},\ \bibinfo {pages}
  {5476--5482} (\bibinfo {year} {2019})}\BibitemShut {NoStop}%
\bibitem [{\citenamefont {Olson~Reichhardt}\ and\ \citenamefont
  {Reichhardt}(2008)}]{Reichhardt08}%
  \BibitemOpen
  \bibfield  {author} {\bibinfo {author} {\bibfnamefont {C.~J.}\ \bibnamefont
  {Olson~Reichhardt}}\ and\ \bibinfo {author} {\bibfnamefont {C.}~\bibnamefont
  {Reichhardt}},\ }\bibfield  {title} {\enquote {\bibinfo {title} {Viscous
  decoupling transitions for individually dragged particles in systems with
  quenched disorder},}\ }\href {\doibase 10.1103/PhysRevE.78.011402} {\bibfield
   {journal} {\bibinfo  {journal} {Phys. Rev. E}\ }\textbf {\bibinfo {volume}
  {78}},\ \bibinfo {pages} {011402} (\bibinfo {year} {2008})}\BibitemShut
  {NoStop}%
\bibitem [{\citenamefont {Ma}\ \emph {et~al.}(2018)\citenamefont {Ma},
  \citenamefont {Reichhardt},\ and\ \citenamefont {Reichhardt}}]{Ma18}%
  \BibitemOpen
  \bibfield  {author} {\bibinfo {author} {\bibfnamefont {X.}~\bibnamefont
  {Ma}}, \bibinfo {author} {\bibfnamefont {C.~J.~O.}\ \bibnamefont
  {Reichhardt}}, \ and\ \bibinfo {author} {\bibfnamefont {C.}~\bibnamefont
  {Reichhardt}},\ }\bibfield  {title} {\enquote {\bibinfo {title} {Manipulation
  of individual superconducting vortices and stick-slip motion in periodic
  pinning arrays},}\ }\href {\doibase 10.1103/PhysRevB.97.214521} {\bibfield
  {journal} {\bibinfo  {journal} {Phys. Rev. B}\ }\textbf {\bibinfo {volume}
  {97}},\ \bibinfo {pages} {214521} (\bibinfo {year} {2018})}\BibitemShut
  {NoStop}%
\bibitem [{\citenamefont {Ma}\ \emph {et~al.}(2020)\citenamefont {Ma},
  \citenamefont {Reichhardt},\ and\ \citenamefont {Reichhardt}}]{Ma20}%
  \BibitemOpen
  \bibfield  {author} {\bibinfo {author} {\bibfnamefont {X.}~\bibnamefont
  {Ma}}, \bibinfo {author} {\bibfnamefont {C.~J.~O.}\ \bibnamefont
  {Reichhardt}}, \ and\ \bibinfo {author} {\bibfnamefont {C.}~\bibnamefont
  {Reichhardt}},\ }\bibfield  {title} {\enquote {\bibinfo {title} {Braiding
  {M}ajorana fermions and creating quantum logic gates with vortices on a
  periodic pinning structure},}\ }\href {\doibase 10.1103/PhysRevB.101.024514}
  {\bibfield  {journal} {\bibinfo  {journal} {Phys. Rev. B}\ }\textbf {\bibinfo
  {volume} {101}},\ \bibinfo {pages} {024514} (\bibinfo {year}
  {2020})}\BibitemShut {NoStop}%
\bibitem [{\citenamefont {Reichhardt}\ and\ \citenamefont
  {Reichhardt}(2007)}]{Reichhardt07a}%
  \BibitemOpen
  \bibfield  {author} {\bibinfo {author} {\bibfnamefont {C.}~\bibnamefont
  {Reichhardt}}\ and\ \bibinfo {author} {\bibfnamefont {C.~J.~Olson}\
  \bibnamefont {Reichhardt}},\ }\bibfield  {title} {\enquote {\bibinfo {title}
  {Commensurability effects at nonmatching fields for vortices in diluted
  periodic pinning arrays},}\ }\href {\doibase 10.1103/PhysRevB.76.094512}
  {\bibfield  {journal} {\bibinfo  {journal} {Phys. Rev. B}\ }\textbf {\bibinfo
  {volume} {76}},\ \bibinfo {pages} {094512} (\bibinfo {year}
  {2007})}\BibitemShut {NoStop}%
\bibitem [{\citenamefont {Reichhardt}\ and\ \citenamefont
  {Olson~Reichhardt}(2009)}]{Reichhardt09}%
  \BibitemOpen
  \bibfield  {author} {\bibinfo {author} {\bibfnamefont {C.}~\bibnamefont
  {Reichhardt}}\ and\ \bibinfo {author} {\bibfnamefont {C.~J.}\ \bibnamefont
  {Olson~Reichhardt}},\ }\bibfield  {title} {\enquote {\bibinfo {title}
  {Transport anisotropy as a probe of the interstitial vortex state in
  superconductors with artificial pinning arrays},}\ }\href {\doibase
  10.1103/PhysRevB.79.134501} {\bibfield  {journal} {\bibinfo  {journal} {Phys.
  Rev. B}\ }\textbf {\bibinfo {volume} {79}},\ \bibinfo {pages} {134501}
  (\bibinfo {year} {2009})}\BibitemShut {NoStop}%
\bibitem [{\citenamefont {Poccia}\ \emph {et~al.}(2015)\citenamefont {Poccia},
  \citenamefont {Baturina}, \citenamefont {Coneri}, \citenamefont {Molenaar},
  \citenamefont {Wang}, \citenamefont {Bianconi}, \citenamefont {Brinkman},
  \citenamefont {Hilgenkamp}, \citenamefont {Golubov},\ and\ \citenamefont
  {Vinokur}}]{Poccia15}%
  \BibitemOpen
  \bibfield  {author} {\bibinfo {author} {\bibfnamefont {N.}~\bibnamefont
  {Poccia}}, \bibinfo {author} {\bibfnamefont {T.~I.}\ \bibnamefont
  {Baturina}}, \bibinfo {author} {\bibfnamefont {F.}~\bibnamefont {Coneri}},
  \bibinfo {author} {\bibfnamefont {C.~G.}\ \bibnamefont {Molenaar}}, \bibinfo
  {author} {\bibfnamefont {X.~R.}\ \bibnamefont {Wang}}, \bibinfo {author}
  {\bibfnamefont {G.}~\bibnamefont {Bianconi}}, \bibinfo {author}
  {\bibfnamefont {A.}~\bibnamefont {Brinkman}}, \bibinfo {author}
  {\bibfnamefont {H.}~\bibnamefont {Hilgenkamp}}, \bibinfo {author}
  {\bibfnamefont {A.~A.}\ \bibnamefont {Golubov}}, \ and\ \bibinfo {author}
  {\bibfnamefont {V.~M.}\ \bibnamefont {Vinokur}},\ }\bibfield  {title}
  {\enquote {\bibinfo {title} {Critical behavior at a dynamic vortex
  insulator-to-metal transition},}\ }\href {\doibase 10.1126/science.1260507}
  {\bibfield  {journal} {\bibinfo  {journal} {Science}\ }\textbf {\bibinfo
  {volume} {349}},\ \bibinfo {pages} {1202--1205} (\bibinfo {year}
  {2015})}\BibitemShut {NoStop}%
\bibitem [{\citenamefont {Jiang}\ \emph {et~al.}(2004)\citenamefont {Jiang},
  \citenamefont {Dikin}, \citenamefont {Chandrasekhar}, \citenamefont
  {Metlushko},\ and\ \citenamefont {Moshchalkov}}]{Jiang04}%
  \BibitemOpen
  \bibfield  {author} {\bibinfo {author} {\bibfnamefont {Z.}~\bibnamefont
  {Jiang}}, \bibinfo {author} {\bibfnamefont {D.~A.}\ \bibnamefont {Dikin}},
  \bibinfo {author} {\bibfnamefont {V.}~\bibnamefont {Chandrasekhar}}, \bibinfo
  {author} {\bibfnamefont {V.~V.}\ \bibnamefont {Metlushko}}, \ and\ \bibinfo
  {author} {\bibfnamefont {V.~V.}\ \bibnamefont {Moshchalkov}},\ }\bibfield
  {title} {\enquote {\bibinfo {title} {Pinning phenomena in a superconducting
  film with a square lattice of artificial pinning centers},}\ }\href {\doibase
  10.1063/1.1767278} {\bibfield  {journal} {\bibinfo  {journal} {Appl. Phys.
  Lett.}\ }\textbf {\bibinfo {volume} {84}},\ \bibinfo {pages} {5371--5373}
  (\bibinfo {year} {2004})}\BibitemShut {NoStop}%
\bibitem [{\citenamefont {Reichhardt}\ and\ \citenamefont
  {Nori}(1999)}]{Reichhardt99}%
  \BibitemOpen
  \bibfield  {author} {\bibinfo {author} {\bibfnamefont {C.}~\bibnamefont
  {Reichhardt}}\ and\ \bibinfo {author} {\bibfnamefont {F.}~\bibnamefont
  {Nori}},\ }\bibfield  {title} {\enquote {\bibinfo {title} {Phase locking,
  devil's staircases, {F}arey trees, and {A}rnold tongues in driven vortex
  lattices with periodic pinning},}\ }\href {\doibase
  10.1103/PhysRevLett.82.414} {\bibfield  {journal} {\bibinfo  {journal} {Phys.
  Rev. Lett.}\ }\textbf {\bibinfo {volume} {82}},\ \bibinfo {pages} {414--417}
  (\bibinfo {year} {1999})}\BibitemShut {NoStop}%
\bibitem [{\citenamefont {Korda}\ \emph {et~al.}(2002)\citenamefont {Korda},
  \citenamefont {Taylor},\ and\ \citenamefont {Grier}}]{Korda02}%
  \BibitemOpen
  \bibfield  {author} {\bibinfo {author} {\bibfnamefont {P.~T.}\ \bibnamefont
  {Korda}}, \bibinfo {author} {\bibfnamefont {M.~B.}\ \bibnamefont {Taylor}}, \
  and\ \bibinfo {author} {\bibfnamefont {D.~G.}\ \bibnamefont {Grier}},\
  }\bibfield  {title} {\enquote {\bibinfo {title} {Kinetically locked-in
  colloidal transport in an array of optical tweezers},}\ }\href {\doibase
  10.1103/PhysRevLett.89.128301} {\bibfield  {journal} {\bibinfo  {journal}
  {Phys. Rev. Lett.}\ }\textbf {\bibinfo {volume} {89}},\ \bibinfo {pages}
  {128301} (\bibinfo {year} {2002})}\BibitemShut {NoStop}%
\end{thebibliography}%

\end{document}